\documentclass{article}

\usepackage{emulateapj}
\usepackage{onecolfloat}
\usepackage{graphicx} 
\usepackage{fancyheadings} 
\usepackage{ulem}
\usepackage{rotating}
\usepackage{lscape}
 
\def\ndla{19~}

\def\delv{\Delta v}

\def\lya{Ly$\alpha$ }

\def\kms{km~s$^{-1}$ }
\def\cm#1{\, {\rm cm^{#1}}}
\def\N#1{{N({\rm #1})}}

\def\rAA{{\rm \, \AA}}
\def\sci#1{{\rm \; \times \; 10^{#1}}}

\def\cmma{\;\;\; ,}

\def\mkms{{\rm \; km\;s^{-1}}}
\newcommand{\tskip}{\tablevspace{3pt}}

\begin{document}

\twocolumn[%
\submitted{ApJ: Accepted October 20, 1998}

\title{CHEMICAL ABUNDANCES OF THE DAMPED \lya SYSTEMS AT $z > 1.5$}

\author{ JASON X. PROCHASKA\altaffilmark{1} 
\& ARTHUR M. WOLFE\altaffilmark{1}\\
Department of Physics, and Center for Astrophysics and Space Sciences; \\
University of California, San
Diego; \\
C--0424; La Jolla; CA 92093\\}

\begin{abstract} 

We present chemical abundance measurements for \ndla damped \lya
systems observed with HIRES on the 10m W.M. Keck Telescope.  
We perform a detailed analysis of every system, deriving ionic column
densities for all unblended metal-line transitions. 
Our principal goal is to investigate the abundance patterns of the damped
systems and thereby determine the underlying
physical processes which dominate their chemical evolution. 
We place particular emphasis on gauging
the relative importance of
two complementary effects often invoked to explain the damped \lya
abundances:  (1) nucleosynthetic enrichment from Type II supernovae
and (2) an ISM-like dust depletion pattern.

Similar to the principal results of Lu et al.\ (1996),  
our observations lend support both for dust 
depletion and Type II SN enrichment. 
Specifically, the observed
overabundance of Zn/Fe and underabundance of Ni/Fe relative to
solar abundances suggest significant
dust depletion within the damped \lya systems.  Meanwhile,
the relative abundances
of Al, Si, and Cr vs.\ Fe are consistent with both dust depletion
and Type II supernova enrichment.
Our measurements of Ti/Fe and the Mn/Fe measurements from Lu et al.\ (1996),
however, cannot be explained by dust depletion and indicate an
underlying Type II SN pattern. 
Finally, the observed values of 
$\lbrack$S/Fe$\rbrack$ are inconsistent with the combined effects of dust
depletion and the nucleosynthetic yields expected for Type II supernovae.
This last result emphasizes the need for another
physical process to explain the damped \lya abundance patterns.

We also examine the metallicity of the damped \lya systems both
with respect to Zn/H and Fe/H.  Our results confirm 
previous surveys by Pettini and collaborators, i.e., 
$\lbrack <$Zn/H$> \rbrack $ = $-1.15 \pm 0.15$~dex.   In contrast with other
damped \lya surveys at $z>1.5$, we do not formally observe 
an evolution of metallicity with redshift,
although we stress this result is based on the statistics
from a small sample of high $z$ damped systems.
\end{abstract}

]

\altaffiltext{1}{Visiting Astronomer, W.M. Keck Telescope.
The Keck Observatory is a joint facility of the University
of California and the California Institute of Technology.}

\pagestyle{fancyplain}
\lhead[\fancyplain{}{\thepage}]{\fancyplain{}{PROCHASKA \& WOLFE}}
\rhead[\fancyplain{}{CHEMICAL ABUNDANCES OF $z>1.5$ DAMPED LYA SYSTEMS}]{\fancyplain{}{\thepage}}
\setlength{\headrulewidth=0pt}
\cfoot{}

\section{INTRODUCTION}

The damped \lya systems dominate the neutral gas content of the
universe and at high redshift are widely believed to be the progenitors
of present-day galaxies (\cite{wol95}).
Therefore, one can directly
measure the chemical evolution of the early universe
by tracing the chemical abundances of the damped systems. 
Pettini and his collaborators have performed the most
extensive surveys on the metallicity of the damped \lya
systems to date (\cite{ptt94,ptt97}).  Working on the 
premise one can measure accurate column
densities of Zn$^+$ and Cr$^+$ from unresolved line profiles,
they have successfully observed over 30 damped \lya systems
with the Anglo-Australian, William Herschel, and Hale
Telescopes. Their results indicate a mean metallicity of
Zn/H $\approx$ 1/10 solar abundance with a notably large dispersion.  
These measurements indicate the damped systems are chemically
young at $z > 2$ and lend further support to the interpretation
of damped systems as the progenitors of modern galaxies
(e.g.\ \cite{mny96}).
In addition to an analysis of Zn, Pettini et al.\ performed 
accurate measurements of $\N{Cr^+}$ aided by the coincidence
in wavelength of the strongest metal-line transitions for the two
species.  Comparing the relative 
abundances of Cr and Zn, they noted
an underabundance of Cr to Zn relative to solar abundances (a typical 
value\footnote{[X/Y] $\equiv {\rm 
\log[X/Y] - \log[X/Y]_\odot}$} is
[Cr/Zn]~$\approx -0.5$~dex).  
In the ISM, Cr is significantly depleted onto dust
grains while Zn is only lightly depleted.  Therefore, Pettini
and others have argued that the relative underabundance of Cr to Zn
is indicative of dust in the damped \lya systems.  
They also point out the Cr/Zn measurements imply a dust-to-gas ratio
much lower than that observed in dusty ISM regions where typical
values for [Cr/Zn] are less than $-1.0$~dex.  
Establishing the level of dust depletion in damped \lya systems
is very important as dust could significantly bias the results
from the damped \lya surveys against high $\N{HI}$ systems (\cite{fal93}).  

\begin{table*}[ht]
\caption{\centerline
{\sc QSO and Observational Data} \smallskip \label{jouobs}}
\begin{center}
\begin{tabular}{llccccc}
\tableline
\tableline \tskip
QSO & Alternate Name & Date
& Exposure
& $z_{em}$
& Resolution & SNR \\
& & Time (s) & & (\kms) \\
\tableline \tskip
Q0019$-$15 & BR 0019$-$1522  & F96 & 35000 & 4.528 & 7.5 & 18 \nl
Q0100+13   & PHL 957         & S94 & 11700 & 2.69 & 7.5 & 40  \nl
Q0149+33   & OC 383          & F97 & 17600 & 2.43 & 7.5 & 25  \nl
Q0201+36   & UT 0201+3634    & F94 & 34580 & 2.49 & 7.5 & 35  \nl
Q0347$-$38 & ...             & F96 & 12600 & 3.23  & 7.5 & 33 \nl
Q0458$-$02 & PKS 0458$-$020  & F95 & 28800 & 2.29 & 7.5 & 15  \nl
Q0841+12   & ...             & F97,S98 & 10800 & & 7.5 & 30  \nl
Q0951$-$04 & BR 0951$-$0450  & S97 & 30600 & 4.369 & 7.5 & 13 \nl
Q1215+33   & GC 1215$+$3322  & S94 & 14040 & 2.61 & 7.5 & 20  \nl
Q1331+17   & MC 1331$+$170   & S94 & 36000 & 2.08 & 6.6 & 80  \nl
Q1346$-$03 & BRI 1346$-$0322 & S97 & 31000 & 3.992 & 7.5 & 29 \nl
Q1759+75   & GB1759+7539     & F96 & 10400 & 3.05  & 6.6 & 33 \nl
Q2206$-$19 & ...             & F94 & 25900 & 2.56 & 7.5 & 40  \nl
Q2230+02   & LBQS 2230+0232  & F97 & 18000 & 2.15 & 7.5 & 26  \nl
Q2231$-$00 & LBQS 2230$-$0015& F95 & 14400 & 3.02 & 7.5 & 30  \nl
Q2348$-$14 & ...             & F96 & 9000  & 2.940 & 7.5 & 41 \nl
Q2359$-$02 & UM 196          & F97 & 25000 & 2.31 & 7.5 & 17  \nl
\tskip \tableline
\end{tabular}
\end{center}
\end{table*}

More recently, Lu et al.\ (1996)
have offered an alternate interpretation for the abundance
patterns of the damped systems.  
Unlike the Pettini et al.\ sample, Lu et al.\ measured abundances
for a large number of elements including Zn, Cr, Fe, S, N, O, Si, Ni, 
and Mn with HIRES on the 10m W.M. Keck Telescope.
The primary result of their analysis 
is that the damped \lya systems exhibit abundance
patterns typical of nucleosynthetic yields for Type II supernova.  The 
tell-tale signature of Type II SN enrichment
is the overabundance of $\alpha$-process
elements (e.g.\ Si and S) relative to Fe.  Empirical measurements 
of the abundance patterns for Type II SN are commonly derived from 
the metal-poor halo stars (\cite{evd93})
which presumably were primarily enriched by Type II supernova.
As expected for Type II SN abundances, 
Lu et al.\ found an overabundance of Si/Fe in every case and
further noted that the relative abundances of Fe, Cr, Mn, N, O, and
S all match the metal-poor halo star observations.
Furthermore, Lu et al.\ stress the measured ratios of 
Mn/Fe and N/O cannot be explained in terms of dust depletion 
and therefore argue for an underlying Type II SN abundance pattern.
The interpretation of the damped \lya abundances patterns 
as Type II SN enrichment fails, however, with respect to Zn.
In particular, the observed overabundance of Zn to Fe (or Cr)
contradicts the
observations of halo stars where one finds [Zn/Fe] $\approx 0$~dex irrespective
of the star's metallicity.  While it
is possible to theoretically explain the observed overabundance of
Zn relative to Fe 
as a natural consequence of Type II supernova enrichment (\cite{hff96}),
the halo star observations pose a significant 
problem.  Several authors have interpreted the observed
abundance patterns by combining the effects of dust depletion
and Type II SN enrichment (\cite{lu96b,kulk97,vld98}), but   
their efforts have been largely unsuccessful. 
They have had particular difficulty in matching both the [Zn/Fe]
and [Mn/Fe] patterns observed in the damped systems.  Developing a consistent
explanation for all of the abundance patterns remains an outstanding 
problem. 

In this paper, we investigate these issues with observations of
\ndla damped \lya systems, including two systems 
previously observed by Lu et al.\ (1996). Building
on the abundance results from Prochaska \& Wolfe (1996) and
Prochaska \& Wolfe (1997a),
we derive ionic column densities for all of
the unblended metal-line transitions comprising our damped 
\lya sample.  In turn, we look for abundance patterns similar
to those observed in the Lu et al.\ (1996) sample and also
interpret these results in the light of dust depletion.  
Lastly, we investigate the evolution of the observed
metallicity of the damped \lya systems with increasing redshift.

In $\S$~\ref{sec-obs}
we summarize the observational sample and data reduction
techniques.  The individual damped systems are briefly discussed
in $\S$~\ref{sec-ion} and
measurements of the ionic column densities and velocity plots
of the metal-line transitions are presented.
$\S$~\ref{sec-abnd} discusses the observed abundance patterns
of the damped \lya systems.  Finally, the metallicity of the
damped \lya systems is investigated in $\S$~\ref{sec-met} and
a brief summary is given in $\S$~\ref{sec-sum}.

\begin{table*} 
\caption{\centerline
{\sc Metal Line Data} \smallskip \label{fosc}}
\begin{center}
\begin{tabular}{lll}
\tableline
\tableline \tskip
Transition & $\lambda_{\rm rest}$ ($\AA$) &
$f^a$ \\
\tableline \tskip
HI 1215 & 1215.6701 & 0.4164 \nl
OI 1302 & 1302.1685 & 0.04887 \nl
SiII 1304 & 1304.3702 & 0.0940 \nl
NiII 1317 & 1317.217  & 0.1458$^b$ \nl
CII 1334 & 1334.5323 & 0.1278 \nl
CuII 1358 & 1358.773 & 0.3803 \nl
NiII 1370 & 1370.131 & 0.144 \nl
SiIV 1393 & 1393.755 & 0.528 \nl
SnII 1400 & 1400.400 & 0.71 \nl
SiIV 1402 & 1402.770 & 0.262 \nl
GaII 1414 & 1414.402 & 1.8 \nl
NiII 1454 & 1454.842 & 0.0516 \nl
SiII 1526 & 1526.7066 & 0.1160 \nl
CIV 1548 & 1548.195 & 0.1908 \nl
CIV 1550 & 1550.770 & 0.09522 \nl
GeII 1602 & 1602.4863 & 0.135 \nl
FeII 1608 & 1608.4511 & 0.06196 \nl
FeII 1611 & 1611.2005 & 0.001020 \nl
AlII 1670 & 1670.7874 & 1.88 \nl
PbII 1682 & 1682.15 & 0.156 \nl
NiII 1703 & 1703.405 & 0.01224 \nl
NiII 1709 & 1709.600 & 0.0666$^b$ \nl
NiII 1741 & 1741.549 & 0.0776$^b$ \nl
NiII 1751 & 1751.910 & 0.0638 \nl
SiII 1808 & 1808.0126 & 0.00218 \nl
AlIII 1854 & 1854.716 & 0.539 \nl
AlIII 1862 & 1862.790 & 0.268 \nl
TiII 1910a & 1910.6 & 0.0975 \nl
TiII 1910b & 1910.97 & 0.0706 \nl
ZnII 2026 & 2026.136 & 0.489 \nl
CrII 2056 & 2056.254 & 0.1050$^c$ \nl
CrII 2062 & 2062.234 & 0.0780$^c$ \nl
ZnII 2062 & 2062.664 & 0.256 \nl
CrII 2066 & 2066.161 & 0.05150$^c$ \nl
FeII 2260 & 2260.7805 & 0.00244 \nl
FeII 2344 &  2344.214 & 0.1108 \nl
FeII 2374 & 2374.4612 & 0.03260 \nl
FeII 2382 & 2382.765 &  0.3006 \nl
MnII 2576 & 2576.877 & 0.3508 \nl
FeII 2586 & 2586.6500 & 0.0684 \nl
MnII 2594 & 2594.499 & 0.2710 \nl
FeII 2600 & 2600.1729 & 0.2132 \nl
MnII 2606 & 2606.462 & 0.1927 \nl
\tskip \tableline
\end{tabular}
\end{center}

\centerline{$^a$Unless otherwise indicated, the $f$ and 
$\lambda$ values were taken from Morton (1991)}
\centerline{$^b$Zsargo \& Federman (1998)}
\centerline{$^c$Tripp et al.\ (1996)}
\end{table*}

\section{OBSERVATIONAL SAMPLE}
\label{sec-obs}

Table~\ref{jouobs} presents a journal of our observations.
In addition to exposure times and dates,
we estimate the typical signal-to-noise ratio per pixel
(SNR) and resolution of the
spectra, and include the emission redshift, $z_{em}$, of the quasar.
All of the data were acquired with the high-resolution
echelle spectrograph (HIRES; \cite{vgt92}) on the 10m W.M. Keck I telescope.
The data were reduced
with the HIRES software package developed by T.\ Barlow.  
This package converts the 2D echelle images to fully
reduced, 1D wavelength-calibrated spectra.  We then continuum fit these
spectra with a program similar to the IRAF package {\it continuum}
and optimally coadded multiple observations.  

Our observational sample of QSO's all have 
at least one
known intervening damped \lya system.  The systems exhibit
a range of $\N{HI}$ values and absorption redshifts ($z_{abs} = 1.8 - 4.2$).
In several cases, we have identified additional
metal-line systems with very large
ionic column densities which suggest they are damped systems.  
In the following, however, we restrict our analysis to systems with
measured HI column density,  $\N{HI} > 2 \sci{20} \cm{-2}$.
The metal-line transitions are identified by composing velocity
plots of the absorption lines listed in Table~\ref{fosc} at the known
redshift of the damped system and then correlating the profiles by eye.  
We performed a systematic search for other metal-line systems toward
each QSO to account for possible line misidentification and blending.
The data are presented in the following section.

\section{IONIC COLUMN DENSITIES}
\label{sec-ion}

All of the ionic column densities presented in this section
were derived with the apparent optical depth method 
(AODM; \cite{sav91}).  
Savage and Sembach (1991) have stressed measuring
column densities by fitting multiple Voigt profiles to
the line-profiles does not always account for
hidden saturated components.  
They introduced a technique to correct for hidden saturation by 
comparing the apparent column density, $N_a$, for multiple transitions
from a single ion.
The analysis involves calculating $N_a(v)$
for each pixel from the optical depth equation
 
\begin{equation}
N_a(v) = {m_e c \over \pi e^2} {\tau_a(v) \over f \lambda} ,
\end{equation}
 
\noindent where $\tau_a(v) = \ln [I_i (v) / I_a (v)]$, f is
the oscillator strength, $\lambda$ is the rest wavelength and
$I_i$ and $I_a$ are the incident and measured intensity.  Comparing $N_a (v)$ 
deduced from two or more transitions of the same ion, one finds the stronger
transition will have smaller values of $N_a (v)$ in those features where
hidden
saturation is present. Thus, one can ascertain the likelihood of saturated
components for ions with multiple transitions.

In Wolfe et al.\ (1994), Prochaska \& Wolfe (1996) 
and Prochaska \& Wolfe (1997a), we 
showed the damped \lya profiles are not contaminated by hidden
saturation. Furthermore, we
demonstrated the column densities derived with
the AODM agree very well with
line-profile fitting, which should give
a more accurate measure of the ionic column densities when
hidden saturation is negligible.
As the AODM is easier to apply to a large data set, we have chosen
to use this technique to measure the ionic column densities for the
damped \lya sample.  Throughout the paper we adopt the wavelengths
and oscillator strengths presented in Table~\ref{fosc} compiled by
Morton (1991), Tripp et al.\ (1996), and Zsarg$\rm \acute o$ 
\& Federman (1998).

Tables~\ref{Q0000-clm}$-$\ref{Q2359B-clm} present the results
of the abundance measurements
including an estimate of the $1\sigma$ error.
For those transitions where the profile saturates (i.e.\ $I_i/I_a < 0.01$
in at least one pixel),
the column densities are listed as lower 
limits.
The values reported as upper limits are $3 \sigma$ upper limits.
We warn the reader of two points: (1) logarithmic errors
are misleadingly small (e.g.\ a 0.1~dex error is $\approx 25\%$ for 
a 13~dex measurement)
and  (2) we have ignored continuum
error in our analysis which could significantly affect measurements
of very weak transitions.
In the following subsections 
we comment briefly on each of the damped \lya systems,
plot all of the identified metal-line
transitions, and 
discuss the adopted $\N{HI}$ values.
We note which systems are members of the LBQS sample (\cite{wol95})
and advise the reader the refer to that paper for further details.
In the velocity plots, $v=0$ is chosen arbitrarily and
corresponds to the redshift listed in the figure caption. 
We indicate regions where blends with other transitions occur (primarily
through blends with other metal-line systems
or the \lya forest) by plotting with dotted lines.

\subsection{Q0000$-$26, $z$ = 3.390}

\begin{figure*}
\begin{center}
\includegraphics[height=8.0in, width=5.0in, bb = 55 48 557 744]{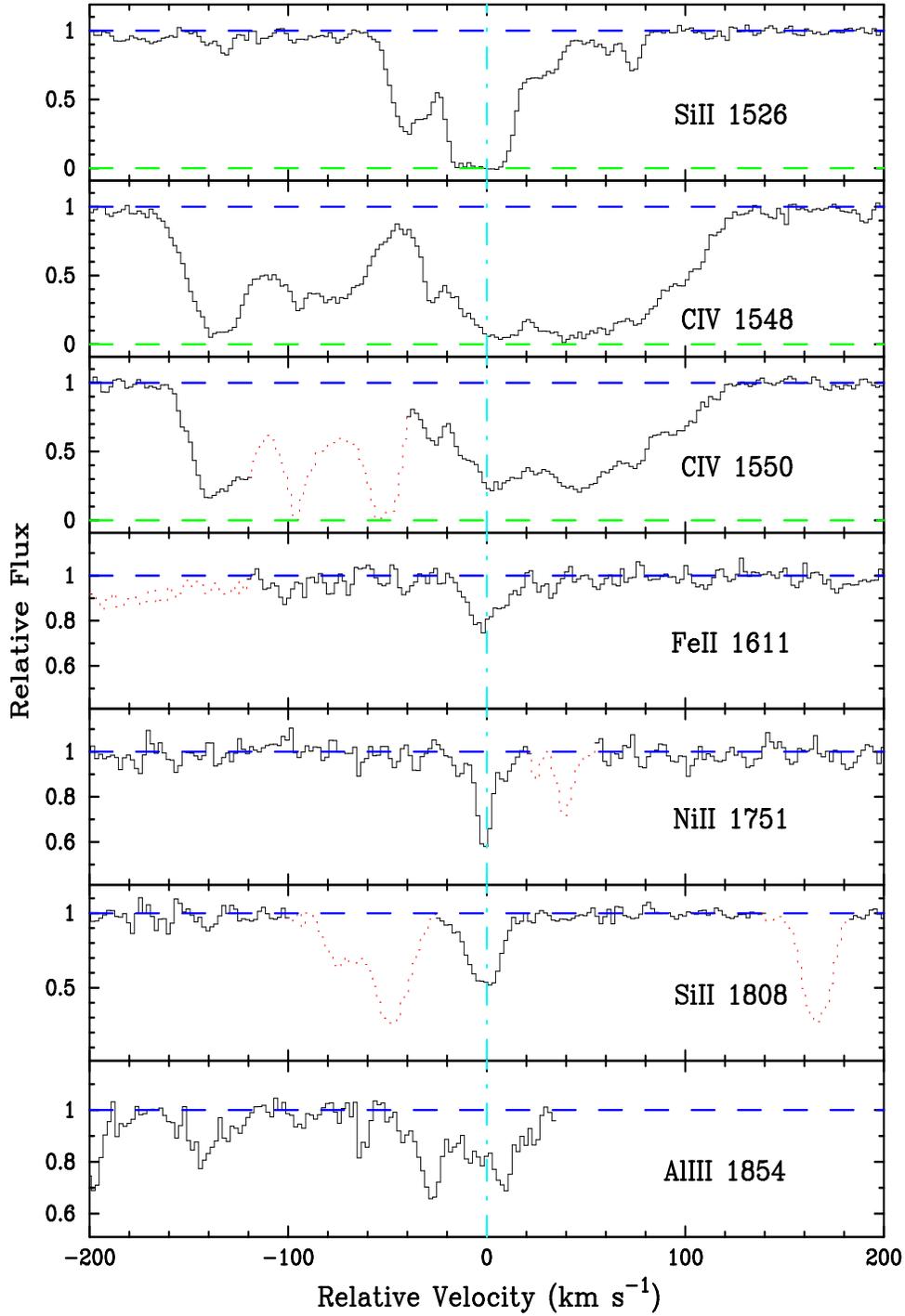}
\caption{Velocity plot of the metal-line transitions for the 
damped \lya system at $z = 3.390$ toward Q0000$-$26.
The vertical line at $v=0$ corresponds to $z = 3.3901$.  Figures 2-21
are not included here but can be downloaded from: 
http://mamacass.ucsd.edu:8080/people/xavier/WWW/DLA/abundfig.ps.gz }
\label{vpQ0000}
\end{center}
\end{figure*}

This high redshift damped \lya system has been previously
observed with HIRES (\cite{lu96b}) and 
we adopt $\log \N{HI} = 21.41 \pm 0.08$ based on that analysis. 
It is a member of the LBQS (\cite{wol95}) statistical sample, one
of the few with $z_{abs} > 3$
The velocity profiles are presented in Figure~\ref{vpQ0000}
and the derived ionic column densities are listed in Table~\ref{Q0000-clm}.
As our spectral coverage did not
include FeII~1608, the Fe abundance is based solely on FeII~1611
which is a very weak transition and has a relatively low SNR.  
Interestingly, we find $\N{Fe^+}$ to be significantly higher than the lower
limit derived from the FeII~1608 transition
reported by Lu et al.\ (1996), which may be the result
of an error in the continuum fit to our profile. 
We find $\N{Si^+}$ to be significantly higher as well, however,
both from the direct
measurement of the SiII~1808 transition and by fitting a Voigt
profile to the saturated SiII~1526 profile.
We therefore note 
Lu et al.\ (1996) may have significantly underestimated the true
metallicity.  At the same time, we observe that the $\N{Ni^+}$
measurement coincides with the Lu et al.\ results.
Because we have no unblended, unsaturated high SNR profiles for this
system we have not included it in the kinematic analyses thus far
(\cite{pro97b,wol98,pro98}).

\begin{table}[ht] \footnotesize
\begin{center}
\caption{\label{Q0000-clm}}
{\sc Ionic Column Densities: Q0000$-$26, $z$ = 3.390 \smallskip} 
\begin{tabular}{lcccc}
\tableline
\tableline \tskip
Ion & $\lambda$ & AODM & $N_{adopt}$ & [X/H] \\
\tableline \tskip
 HI & 1215 & $21.410 \pm 0.080$ &  \nl
 CIV & 1548 & $14.707 \pm 0.006$ &  \nl
 AlIII & 1854 & $12.772 \pm 0.020$ &  \nl
 SiII & 1526 & $ > 14.607$ & $15.086 \pm 0.012$ & $-1.874 \pm 0.081$ \nl
 & 1808 & $15.086 \pm 0.012$ &  \nl
 FeII & 1611 & $15.146 \pm 0.037$ & $15.146 \pm 0.037$ & $-1.774 \pm 0.088$ \nl
 NiII & 1751 & $13.325 \pm 0.029$ & $13.325 \pm 0.029$ & $-2.335 \pm 0.085$ \nl
\tskip \tableline
\end{tabular}
\end{center}
\end{table}

\subsection{Q0019$-$15, $z$ = 3.439}

This damped \lya system comes from the high redshift survey by
Storrie-Lombardi et al.\ (1996) and the adopted 
$\log \N{HI} = 20.9 \pm 0.1$ was taken from recent Keck measurements
of Storrie-Lombardi \& Wolfe (1998).  
While our observations covered \lya for this 
system, the profile extends over two echelle orders and an accurate measurement
of $\N{HI}$ proved impossible.
Figure~\ref{vpQ0019} shows the metal-line transitions and
Table~\ref{Q0019-clm} presents the measurements for this system.
The Fe abundance is based on the marginally
saturated FeII 1608 profile, yet should be reasonably accurate.
The NiII lines are very weak and in poor SNR regions so 
these measurements are not reliable.  The same is true for the Si
measurements, although to a lesser extent. 
Note all of the high-ion profiles are blended with other metal-line
transitions or \lya forest clouds. 

\begin{table}[ht] \footnotesize 
\begin{center}
\caption{ \label{Q0019-clm}}
{\sc Ionic Column Densities: Q0019$-$15, $z$ = 3.439} 
\begin{tabular}{lcccc}
\tableline
\tableline \tskip
Ion & $\lambda$ & AODM & $N_{adopt}$ & [X/H] \\
\tableline \tskip
 HI & 1215 & $20.900 \pm 0.100$ &  \nl
 SiII & 1526 & $ > 14.953$ & $15.423 \pm 0.053$ & $-1.027 \pm 0.113$ \nl
 & 1808 & $15.423 \pm 0.053$ &  \nl
 SiIV & 1402 & $ > 14.495$ &  \nl
 FeII & 1608 & $14.770 \pm 0.064$ & $14.770 \pm 0.064$ & $-1.640 \pm 0.119$ \nl
 NiII & 1709 & $13.300 \pm 0.103$ & $13.442 \pm 0.043$ & $-1.708 \pm 0.109$ \nl
 & 1741 & $13.442 \pm 0.043$ &  \nl
\tskip \tableline
\end{tabular}
\end{center}
\end{table}

\subsection{Q0100$+$13, $z$ = 2.309}

The majority of our results on PHL 957 (a.k.a.\ Q0100$+$13)
were published by Wolfe et al.\ (1994).  
The major exception is we now present
a measurement of the Fe abundance based on the previously unidentified
FeII~1611 profile.  
Also, we measure column densities in light of new $f$ and $\lambda$
values, in particular those for the CrII and NiII transitions.
Note the O abundance is based on the very
weak OI~1355 profile and we report it as a $3\sigma$ upper limit.
Figure~\ref{vpQ0100} presents the velocity plots and Table~\ref{Q0100-clm}
lists the measured ionic column densities for this system.
This damped system is included in the statistical sample of the 
LBQS survey.

\begin{table}[hb] \footnotesize
\begin{center}
\caption{ \label{Q0100-clm}}
{\sc Ionic Column Densities: Q0100$+$13, $z$ = 2.309} 
\begin{tabular}{lcccc}
\tableline
\tableline \tskip
Ion & $\lambda$ & AODM & $N_{adopt}$ & [X/H] \\
\tableline \tskip
 HI & 1215 & $21.400 \pm 0.050$ &  \nl
 CIV & 1548 & $13.241 \pm 0.031$ &  \nl
 & 1550 & $13.303 \pm 0.056$ &  \nl
 OI & 1355 & $ < 17.628$ & $< 17.628$ & $< -0.702$ \nl
 AlIII & 1854 & $12.635 \pm 0.022$ &  \nl
 & 1862 & $12.715 \pm 0.033$ &  \nl
 SiII & 1526 & $ > 14.722$ & $> 14.722$ & $> -2.228$ \nl
 SiIV & 1393 & $13.127 \pm 0.017$ &  \nl
 & 1402 & $13.146 \pm 0.029$ &  \nl
 CrII & 2062 & $13.389 \pm 0.018$ & $13.387 \pm 0.015$ & $-1.693 \pm 0.052$ \nl
 & 2066 & $13.383 \pm 0.024$ &  \nl
 FeII & 1608 & $ > 14.599$ & $15.096 \pm 0.041$ & $-1.814 \pm 0.065$ \nl
 & 1611 & $15.096 \pm 0.041$ &  \nl
 NiII & 1454 & $13.621 \pm 0.041$ & $13.620 \pm 0.014$ & $-2.030 \pm 0.052$ \nl
 & 1741 & $13.620 \pm 0.015$ &  \nl
 ZnII & 2026 & $12.498 \pm 0.028$ & $12.494 \pm 0.023$ & $-1.556 \pm 0.055$ \nl
 & 2062 & $12.485 \pm 0.039$ &  \nl
\tskip \tableline
\end{tabular}
\end{center}
\end{table}

\subsection{Q0149$+$33, $z$ = 2.140}

This relatively metal-poor ([Fe/H]~$\approx -1.8$~dex)
damped system is a also member of the LBQS statistical sample. 
Table~\ref{Q0149-clm} gives the measured column densities and 
Figure~\ref{vpQ0149} plots the metal-lines.
In the following analysis,
we assume $\log \N{HI} = 20.5 \pm 0.1$ (\cite{wol95}).
As with PH957, the O abundance is based on the statistically insignificant
OI~1355 profile and provides a very conservative upper limit.
Finally the Al abundance is derived from the marginally saturated
AlII~1670 profile but should be a reliable value. 
The system is notable for exhibiting an atypical 
Cr to Zn ratio; [Cr/Zn]~$= +0.26 \pm 0.1$~dex.
The Cr measurement is reasonably accurate and the 
ZnII~2062 transition places a rather strict upper limit to the
Zn abundance.  Therefore, it is very likely [Cr/Zn]~$> 0$
which marks the first such occurrence in a damped system and indicates
this system must be essentially undepleted.

\begin{table}[ht] \footnotesize
\begin{center}
\caption{ \label{Q0149-clm}}
{\sc Ionic Column Densities: Q0149$+$33, $z$ = 2.140} 
\begin{tabular}{lcccc}
\tableline
\tableline \tskip
Ion & $\lambda$ & AODM & $N_{adopt}$ & [X/H] \\
\tableline \tskip
 HI & 1215 & $20.500 \pm 0.100$ &  \nl
 CI & 1560 & $ < 12.801$ &  \nl
 CII & 1334 & $ > 14.623$ &  \nl
 & 1335 & $ < 12.780$ &  \nl
 CIV & 1548 & $13.880 \pm 0.011$ &  \nl
 & 1550 & $13.890 \pm 0.017$ &  \nl
 OI & 1355 & $ < 17.795$ & $< 17.795$ & $<  0.365$ \nl
 AlII & 1670 & $12.936 \pm 0.018$ & $12.936 \pm 0.018$ & $-2.044 \pm 0.102$ \nl
 AlIII & 1854 & $12.518 \pm 0.028$ &  \nl
 & 1862 & $12.556 \pm 0.043$ &  \nl
 SiII & 1304 & $14.401 \pm 0.042$ & $14.367 \pm 0.029$ & $-1.683 \pm 0.104$ \nl
 & 1526 & $14.320 \pm 0.034$ &  \nl
 & 1808 & $14.572 \pm 0.047$ &  \nl
 SiIV & 1393 & $13.380 \pm 0.017$ &  \nl
 CrII & 2056 & $12.793 \pm 0.044$ & $12.811 \pm 0.036$ & $-1.369 \pm 0.106$ \nl
 & 2062 & $12.533 \pm 0.087$ &  \nl
 & 2066 & $12.853 \pm 0.060$ &  \nl
 FeII & 1608 & $14.202 \pm 0.018$ & $14.202 \pm 0.018$ & $-1.808 \pm 0.102$ \nl
 & 1611 & $ < 14.785$ &  \nl
 NiII & 1370 & $12.920 \pm 0.103$ & $12.900 \pm 0.042$ & $-1.850 \pm 0.108$ \nl
 & 1703 & $ < 13.575$ &  \nl
 & 1709 & $12.871 \pm 0.088$ &  \nl
 & 1741 & $12.990 \pm 0.064$ &  \nl
 & 1751 & $12.817 \pm 0.090$ &  \nl
 ZnII & 2026 & $11.496 \pm 0.103$ & $11.496 \pm 0.103$ & $-1.654 \pm 0.144$ \nl
 & 2062 & $ < 11.697$ &  \nl
\tskip \tableline
\end{tabular}
\end{center}
\end{table}

\subsection{Q0347$-$38, $z$ = 3.025}

The damped \lya system toward
Q0347$-$38 is another member of the LBQS statistical sample, 
one of the four with $z_{abs} > 3$.
We adopt $\log \N{HI} = 20.8 \pm 0.1$ based on a measurement
by Pettini et al.\ (1994).
This is one of the few systems where we have an estimate of
$\N{S^+}$, although we note SII~1259 is partially
blended with the \lya forest and should be considered an 
upper limit to the true S abundance, $\N{S^+} < 10^{14.73} \cm{-2}$.
We discuss the S/Fe ratio in detail below, noting here that
the value has particular impact on interpreting the abundances
of the damped \lya systems with respect to Type II SN enrichment
and dust depletion.  The system is also notable for the easily
identifiable excited fine structure CII$^*$ 1335 transition.
Unfortunately, the CII~1334 profile is so heavily saturated
that no meaningful comparison can be made with the fine
structure transition.  
Finally, we observe a very low Ni abundance
for this system, [Ni/H]~$< -2.37$~dex, implying [Ni/Fe]~$< -0.5$~dex
which may be difficult to explain within the leading explanations
for the damped \lya abundance patterns.

\begin{table}[ht] \footnotesize
\begin{center}
\caption{
\label{Q0347-clm}}
{\sc Ionic Column Densities: Q0347$-$38, $z$ = 3.025} 
\begin{tabular}{lcccc}
\tableline
\tableline \tskip
Ion & $\lambda$ & AODM & $N_{adopt}$ & [X/H] \\
\tableline \tskip
 HI & 1215 & $20.800 \pm 0.100$ &  \nl
 CII & 1334 & $ > 15.027$ & $> 15.027$ & $> -2.333$ \nl
 & 1335 & $13.477 \pm 0.032$ &  \nl
 CIV & 1548 & $13.846 \pm 0.011$ &  \nl
 & 1550 & $13.783 \pm 0.020$ &  \nl
 OI & 1302 & $ > 15.379$ & $< 17.740$ & $<  0.010$ \nl
 & 1355 & $ < 17.740$ &  \nl
 SiII & 1260 & $ > 14.268$ & $14.820 \pm 0.028$ & $-1.530 \pm 0.104$ \nl
 & 1304 & $14.820 \pm 0.028$ &  \nl
 & 1526 & $ > 14.748$ &  \nl
 SiIV & 1393 & $13.750 \pm 0.017$ &  \nl
 SII & 1259 & $14.731 \pm 0.012$ & $14.731 \pm 0.012$ & $-1.339 \pm 0.101$ \nl
 FeII & 1608 & $14.472 \pm 0.007$ & $14.472 \pm 0.007$ & $-1.838 \pm 0.100$ \nl
 & 1611 & $ < 14.476$ &  \nl
 NiII & 1370 & $ < 12.677$ & $< 12.677$ & $< -2.373$ \nl
\tskip \tableline
\end{tabular}
\end{center}
\end{table}

\subsection{Q0458$-$02, $z$ = 2.040}

This damped \lya system is 'famous' for exhibiting HI 21cm absorption.
In particular, Briggs et al.\ (1989) have used VLBI radio observations
to place a lower limit on its size of 8 $h_{100}^{-1}$~kpc.  
In our analysis we adopt $\log \N{HI} = 21.65 \pm 0.09$ taken from
Pettini et al.\ (1994).  
A plot of the metal-line profiles is given in Figure~\ref{vpQ0458}
and Table~\ref{Q0458-clm} lists the column densities.
Note the CII$^*$ 1335 profile is heavily saturated.  
Assuming the $^2P_{3/2}$ excited fine-structure state is populated
by e$^-$ collisions and 
$\N{C^+} < 10^{17} \cm{-2}$ (which follows by assuming 
[C/H] $<$ [Zn/H]), the limit on $\N{CII^*}$ indicates
$n_e > 0.1 \cm{-3}$.  The system, with one of the highest measured
$\N{HI}$ values, must have a high neutral fraction implying
$n_{H} > 0.1 \cm{-3}$.
Because FeII~1608 is saturated, we base the Fe abundance on the
FeII~1611 transition.   Finally, we note
the ZnII~2026 profile is blended with an unidentified
line at $v > 20 \mkms$ and therefore the AODM was applied only to
$v = +20 \mkms$.  

\begin{table}[ht] \footnotesize
\begin{center}
\caption{ \label{Q0458-clm}}
{\sc Ionic Column Densities: Q0458$-$02, $z$ = 3.040} 
\begin{tabular}{lcccc}
\tableline
\tableline \tskip
Ion & $\lambda$ & AODM & $N_{adopt}$ & [X/H] \\
\tableline \tskip
 HI & 1215 & $21.650 \pm 0.090$ &  \nl
 CII & 1334 & $ > 15.010$ &  \nl
 & 1335 & $ > 14.794$ &  \nl
 CIV & 1548 & $ > 14.906$ &  \nl
 & 1550 & $ > 15.111$ &  \nl
 OI & 1302 & $ > 15.410$ &  \nl
 & 1355 & $ < 18.560$ &  \nl
 AlII & 1670 & $ > 13.720$ &  \nl
 AlIII & 1854 & $13.334 \pm 0.018$ &  \nl
 & 1862 & $13.335 \pm 0.022$ &  \nl
 SiII & 1304 & $ > 15.095$ & $> 15.095$ & $> -2.105$ \nl
 & 1526 & $ > 14.981$ &  \nl
 & 1808 & $ > 16.021$ &  \nl
 SiIV & 1393 & $ > 14.262$ &  \nl
 & 1402 & $ > 14.481$ &  \nl
 CrII & 2056 & $13.756 \pm 0.013$ & $13.797 \pm 0.008$ & $-1.533 \pm 0.090$ \nl
 & 2062 & $13.763 \pm 0.014$ &  \nl
 & 2066 & $13.987 \pm 0.015$ &  \nl
 FeII & 1608 & $ > 15.136$ & $15.508 \pm 0.048$ & $-1.652 \pm 0.102$ \nl
 & 1611 & $15.508 \pm 0.048$ &  \nl
 NiII & 1317 & $13.985 \pm 0.024$ & $13.853 \pm 0.019$ & $-2.047 \pm 0.092$ \nl
 & 1709 & $13.840 \pm 0.035$ &  \nl
 & 1741 & $13.936 \pm 0.032$ &  \nl
 & 1751 & $13.808 \pm 0.033$ &  \nl
 ZnII & 2026 & $13.141 \pm 0.018$ & $13.141 \pm 0.018$ & $-1.159 \pm 0.092$ \nl
\tskip \tableline
\end{tabular}
\end{center}
\end{table}

\subsection{Q0841+12, $z$ = 2.375 \& $z$ = 2.476}

The damped \lya systems toward this BL Lac object were first
identified by C.\ Hazard and were subsequently analyzed by
Pettini et al.\ (1997).
We take $\N{HI} = 20.95 \pm 0.087$ 
for the system at $z = 2.375$
and $\N{HI} = 20.78 \pm 0.097$ for the system at $z = 2.476$ (\cite{ptt97}).
Figures~\ref{vpQ0841A} and \ref{vpQ0841B} plot the metal-line profiles
for these systems and Tables~\ref{Q0841A-clm} and \ref{Q0841B-clm}
list the ionic column densities.  The profiles for the lower redshift
system have good SNR and the derived column densities are accurate.
Unfortunately, its FeII~1608 and FeII~1611 transitions are blended
with sky lines.
In the abundance analysis, then,
we will adopt an Fe abundance, [Fe/H] = [Cr/H],
motivated by the fact that [Cr/Fe]~$\approx$~0 in the damped systems.
For the system at $z = 2.476$, SiII~1808 is blended with the
AlIII 1862 transition from the $z=2.375$ system and also 
may be contaminated by a
sky line.  Therefore, we adopt a lower limit to $\N{Si^+}$
from the saturated SiII~1526 profile.
Finally, we obtain an upper limit measurement
for $\N{Zn^+}$ for this system which is a significant
improvement over previous efforts (\cite{ptt97}).

\begin{table}[ht] \footnotesize
\begin{center}
\caption{ \label{Q0841A-clm}}
{\sc Ionic Column Densities: Q0841+12, $z$ = 2.375} 
\begin{tabular}{lcccc}
\tableline
\tableline \tskip
Ion & $\lambda$ & AODM & $N_{adopt}$ & [X/H] \\
\tableline \tskip
 HI & 1215 & $20.950 \pm 0.087$ &  \nl
 CIV & 1548 & $13.968 \pm 0.011$ &  \nl
 & 1550 & $13.973 \pm 0.015$ &  \nl
 AlII & 1670 & $ > 13.267$ & $> 13.267$ & $> -2.163$ \nl
 AlIII & 1854 & $12.504 \pm 0.029$ &  \nl
 & 1862 & $12.703 \pm 0.035$ &  \nl
 SiII & 1526 & $ > 14.566$ & $15.239 \pm 0.024$ & $-1.261 \pm 0.090$ \nl
 & 1808 & $15.239 \pm 0.024$ &  \nl
 CrII & 2056 & $12.960 \pm 0.061$ & $13.079 \pm 0.027$ & $-1.551 \pm 0.091$ \nl
 & 2062 & $13.185 \pm 0.034$ &  \nl
 & 2066 & $13.035 \pm 0.059$ &  \nl
 NiII & 1454 & $13.211 \pm 0.075$ & $13.243 \pm 0.030$ & $-1.957 \pm 0.092$ \nl
 & 1741 & $13.297 \pm 0.039$ &  \nl
 & 1751 & $13.185 \pm 0.055$ &  \nl
 ZnII & 2026 & $12.114 \pm 0.057$ & $12.115 \pm 0.049$ & $-1.485 \pm 0.100$ \nl
 & 2062 & $12.115 \pm 0.099$ &  \nl
\tskip \tableline
\end{tabular}
\end{center}
\end{table}

\begin{table}[ht] \footnotesize
\begin{center}
\caption{ \label{Q0841B-clm}}
{\sc Ionic Column Densities: Q0841$+$12, $z$ = 2.476} 
\begin{tabular}{lcccc}
\tableline
\tableline \tskip
Ion & $\lambda$ & AODM & $N_{adopt}$ & [X/H] \\
\tableline \tskip
 HI & 1215 & $20.780 \pm 0.097$ &  \nl
 CIV & 1548 & $13.914 \pm 0.009$ &  \nl
 & 1550 & $13.986 \pm 0.012$ &  \nl
 AlII & 1670 & $13.186 \pm 0.286$ & $13.186 \pm 0.286$ & $-2.074 \pm 0.302$ \nl
 AlIII & 1854 & $12.683 \pm 0.027$ &  \nl
 & 1862 & $12.633 \pm 0.038$ &  \nl
 SiII & 1526 & $ > 14.461$ & $> 14.461$ & $> -1.869$ \nl
 & 1808 & $14.958 \pm 0.022$ &  \nl
 SiIV & 1393 & $13.646 \pm 0.009$ &  \nl
 & 1402 & $13.646 \pm 0.011$ &  \nl
 TiII & 1910 & $ < 12.434$ & $< 12.434$ & $< -1.276$ \nl
 CrII & 2056 & $12.907 \pm 0.043$ & $12.840 \pm 0.036$ & $-1.620 \pm 0.103$ \nl
 & 2062 & $12.759 \pm 0.061$ &  \nl
 & 2066 & $ < 12.802$ &  \nl
 FeII & 1608 & $14.434 \pm 0.027$ & $14.434 \pm 0.027$ & $-1.856 \pm 0.101$ \nl
 & 1611 & $ < 14.668$ &  \nl
 NiII & 1741 & $13.067 \pm 0.052$ & $13.048 \pm 0.049$ & $-1.982 \pm 0.108$ \nl
 & 1751 & $12.965 \pm 0.129$ &  \nl
 ZnII & 2026 & $ < 11.778$ & $< 11.778$ & $< -1.652$ \nl
 & 2062 & $ < 11.772$ &  \nl
\tskip \tableline
\end{tabular}
\end{center}
\end{table}

\subsection{Q0951$-$04, $z$ = 3.857 \& $z$ = 4.203}

The QSO Q0951$-$04 from the survey by Storrie-Lombardi et al.\ (1996)
has two intervening damped \lya systems, both
at very high redshift.  
The velocity plots and measurements for the $z = 3.857$ system
are given in Figure~\ref{vpQ0951A} and Table~\ref{Q0951A-clm},
while those for the $z = 4.203$ system are presented by Figure~\ref{vpQ0951B}
and Table~\ref{Q0951B-clm}.
We adopt $\log \N{HI} = 20.6 \pm 0.1$ for the
system at $z=3.857$ and $\log \N{HI} = 20.4 \pm 0.1$ for the system
at $z=4.203$, based on recent Keck measurements (\cite{storr98}).
Because all of the 
column densities are based on either marginally saturated profiles
or weaker, low SNR profiles all of these measurements
are somewhat tentative.  
In fact, we consider the limits on $\N{Fe^+}$ for the system at $z = 4.203$
to be too conservative to include this system in the abundance analysis.
Finally, we note the feature at $v = 180 \mkms$
in the NiII 1370 profile for the $z = 3.857$ system may
be an unidentified blend, although it nearly coincides with a 
strong feature in the SiII 1526 profile.

\begin{table}[ht] \footnotesize
\begin{center}
\caption{ \label{Q0951A-clm}}
{\sc Ionic Column Densities: Q0951$-$04, $z$ = 3.857} 
\begin{tabular}{lcccc}
\tableline
\tableline \tskip
Ion & $\lambda$ & AODM & $N_{adopt}$ & [X/H] \\
\tableline \tskip
 HI & 1215 & $20.600 \pm 0.100$ &  \nl
 AlII & 1670 & $13.298 \pm 0.022$ & $13.298 \pm 0.022$ & $-1.782 \pm 0.102$ \nl
 SiII & 1526 & $14.645 \pm 0.030$ & $14.645 \pm 0.030$ & $-1.505 \pm 0.104$ \nl
 SiIV & 1393 & $13.900 \pm 0.011$ &  \nl
 FeII & 1608 & $14.062 \pm 0.060$ & $14.062 \pm 0.060$ & $-2.048 \pm 0.117$ \nl
 NiII & 1370 & $12.994 \pm 0.099$ & $12.994 \pm 0.099$ & $-1.856 \pm 0.141$ \nl
\tskip \tableline
\end{tabular}
\end{center}
\end{table}

\begin{table}[ht] \footnotesize
\begin{center}
\caption{ \label{Q0951B-clm}}
{\sc Ionic Column Densities: Q0951$-$04, $z$ = 4.203} 
\begin{tabular}{lcccc}
\tableline
\tableline \tskip
Ion & $\lambda$ & AODM & $N_{adopt}$ & [X/H] \\
\tableline \tskip
 HI & 1215 & $20.400 \pm 0.100$ &  \nl
 OI & 1302 & $14.607 \pm 0.323$ & $14.607 \pm 0.323$ & $-2.723 \pm 0.339$ \nl
 SiII & 1190 & $13.455 \pm 0.050$ & $13.392 \pm 0.032$ & $-2.558 \pm 0.105$ \nl
 & 1304 & $13.286 \pm 0.060$ &  \nl
 & 1526 & $13.483 \pm 0.055$ &  \nl
 FeII & 1608 & $ < 13.281$ & $< 13.281$ & $< -2.629$ \nl
 & 1611 & $ < 15.153$ &  \nl
 NiII & 1317 & $ < 12.589$ & $< 12.589$ & $< -2.061$ \nl
 & 1370 & $ < 12.625$ &  \nl
\tskip \tableline
\end{tabular}
\end{center}
\end{table}

\subsection{Q1215$+$33, $z$ = 1.999}

This radio-selected damped system (\cite{wol86})
was observed as part of the commissioning
run for the HIRES instrument and is a
member of the LBQS statistical sample.
In the following analysis we assume $\log \N{HI} = 20.95 \pm 0.067$
based on observations by Pettini et al.\ (1994).
Figure~\ref{vpQ1215} presents a plot of the metal-line transitions
and the ionic column densities are listed in Table~\ref{Q1215-clm}.
As the FeII 1608 profile is saturated and the 
FeII 1611 transition is marginally detected, we establish only a lower limit 
to $\N{Fe^+}$.   We will include this system in the abundance
analysis, however, by adopting $\log10[\N{Fe^+}] = 14.648 \pm 0.033$ based
solely on the FeII~1608 profile, noting
this value may be an underestimate.

\begin{table}[ht] \footnotesize
\begin{center}
\caption{ \label{Q1215-clm}}
{\sc Ionic Column Densities: Q1215$+$33, $z$ = 1.999} 
\begin{tabular}{lcccc}
\tableline
\tableline \tskip
Ion & $\lambda$ & AODM & $N_{adopt}$ & [X/H] \\
\tableline \tskip
 HI & 1215 & $20.950 \pm 0.067$ &  \nl
 CIV & 1548 & $13.605 \pm 0.019$ &  \nl
 & 1550 & $13.672 \pm 0.030$ &  \nl
 OI & 1355 & $ < 17.952$ & $< 17.952$ & $<  0.072$ \nl
 AlII & 1670 & $ > 13.358$ &  \nl
 AlIII & 1854 & $12.746 \pm 0.017$ &  \nl
 & 1862 & $12.783 \pm 0.021$ &  \nl
 SiII & 1526 & $ > 14.681$ & $15.030 \pm 0.025$ & $-1.470 \pm 0.072$ \nl
 & 1808 & $15.030 \pm 0.025$ &  \nl
 SiIV & 1393 & $12.993 \pm 0.039$ &  \nl
 CrII & 2056 & $13.173 \pm 0.034$ & $13.130 \pm 0.031$ & $-1.500 \pm 0.074$ \nl
 & 2062 & $13.034 \pm 0.066$ &  \nl
 FeII & 1608 & $14.648 \pm 0.039$ & $14.648 \pm 0.039$ & $-1.812 \pm 0.078$ \nl
 & 1611 & $14.925 \pm 0.100$ &  \nl
 NiII & 1741 & $13.419 \pm 0.040$ & $13.344 \pm 0.033$ & $-1.856 \pm 0.075$ \nl
 & 1751 & $13.262 \pm 0.056$ &  \nl
 ZnII & 2026 & $12.291 \pm 0.058$ & $12.291 \pm 0.058$ & $-1.309 \pm 0.089$ \nl
\tskip \tableline
\end{tabular}
\end{center}
\end{table}

\subsection{Q1331$+$17, $z$ = 1.776}

This famous damped \lya system which exhibits 21cm
absorption (\cite{wol79}) has been studied by a number
of authors (over $100$ papers), 
yet never with the quality of data presented here
(FWHM resolution $\approx 6 \mkms$ and SNR $>$ 50).
Table~\ref{Q1331-clm} lists the column densities for the metal-line
transitions and Figure~\ref{vpQ1331} plots their profiles.
We assume $\log \N{HI} = 21.176 \pm 0.041$ (\cite{ptt94}).
The SNR is excellent throughout 
the entire spectrum yielding very accurate
column density measurements.  
This is one of the very few systems where CI absorption is detected and
Songaila et al.\ (1994) have used measurements of the CI profile
to estimate the cosmic
background temperature at $z = 1.776$.  Consider
the feature at $v = +20 \mkms$ which is fully
resolved in the CI profiles, barely
resolved in the Zn$^+$ profiles, and is unresolved in the stronger
transitions even at 6~\kms resolution.  This suggest the gas in
this component is at a temperature $T < 6000$K.  While
this system may be atypical because it is one of the
few damped systems to exhibit CI absorption, it is worth noting
a majority of observed damped \lya profiles may 
be the result of the superposition of many narrow components.

\begin{table}[ht] \footnotesize
\begin{center}
\caption{ \label{Q1331-clm}}
{\sc Ionic Column Densities: Q1331$+$17, $z$ = 1.776} 
\begin{tabular}{lcccc}
\tableline
\tableline \tskip
Ion & $\lambda$ & AODM & $N_{adopt}$ & [X/H] \\
\tableline \tskip
 HI & 1215 & $21.176 \pm 0.041$ &  \nl
 CI & 1560 & $13.573 \pm 0.013$ &  \nl
 & 1656 & $13.312 \pm 0.012$ &  \nl
 CIV & 1548 & $ > 15.073$ &  \nl
 & 1550 & $ > 15.172$ &  \nl
 AlII & 1670 & $ > 13.573$ &  \nl
 AlIII & 1854 & $13.004 \pm 0.004$ &  \nl
 & 1862 & $12.968 \pm 0.007$ &  \nl
 SiII & 1526 & $ > 14.951$ & $15.285 \pm 0.004$ & $-1.441 \pm 0.041$ \nl
 & 1808 & $15.285 \pm 0.004$ &  \nl
 CrII & 2056 & $12.950 \pm 0.017$ & $12.919 \pm 0.015$ & $-1.937 \pm 0.044$ \nl
 & 2066 & $12.834 \pm 0.034$ &  \nl
 FeII & 1608 & $14.601 \pm 0.003$ & $14.598 \pm 0.001$ & $-2.088 \pm 0.041$ \nl
 & 2344 & $14.597 \pm 0.022$ &  \nl
 & 2374 & $14.598 \pm 0.002$ &  \nl
 & 2382 & $ > 14.433$ &  \nl
 NiII & 1709 & $13.166 \pm 0.017$ & $13.235 \pm 0.009$ & $-2.191 \pm 0.042$ \nl
 & 1741 & $13.313 \pm 0.011$ &  \nl
 & 1751 & $13.165 \pm 0.022$ &  \nl
 ZnII & 2026 & $12.605 \pm 0.009$ & $12.605 \pm 0.008$ & $-1.221 \pm 0.042$ \nl
 & 2062 & $12.605 \pm 0.013$ &  \nl
\tskip \tableline
\end{tabular}
\end{center}
\end{table}

\subsection{Q1346$-$03, $z$ = 3.736}

This very high $z$ damped \lya system was taken from the survey
of Storrie-Lombardi et al.\ (1996).  
We present the velocity plots and column densities in Figure~\ref{vpQ1346}
and Table~\ref{Q1346-clm}. We adopt 
$\log \N{HI} = 20.7 \pm 0.1$ (\cite{storr98}) throughout the analysis.
Unfortunately both FeII 1608 and FeII 1611 are blended with 
B-band sky lines.  Therefore, we have no metallicity
indicator for this system and it is not included in the subsequent analysis.
Measuring [Si/H] = $-2.3$~dex, we expect this is
a very metal-poor system.

\begin{table}[ht] \footnotesize
\begin{center}
\caption{ \label{Q1346-clm}}
{\sc Ionic Column Densities: Q1346$-$03, $z$ = 3.736} 
\begin{tabular}{lcccc}
\tableline
\tableline \tskip
Ion & $\lambda$ & AODM & $N_{adopt}$ & [X/H] \\
\tableline \tskip
 HI & 1215 & $20.700 \pm 0.100$ &  \nl
 CII & 1334 & $14.422 \pm 0.095$ & $14.422 \pm 0.095$ & $-2.838 \pm 0.138$ \nl
 AlII & 1670 & $12.546 \pm 0.025$ & $12.546 \pm 0.025$ & $-2.634 \pm 0.103$ \nl
 SiII & 1304 & $13.961 \pm 0.012$ & $13.954 \pm 0.011$ & $-2.296 \pm 0.101$ \nl
 & 1526 & $13.923 \pm 0.026$ &  \nl
 SiIV & 1393 & $12.344 \pm 0.096$ &  \nl
 & 1402 & $ < 12.598$ &  \nl
 NiII & 1370 & $ < 12.694$ &  \nl
 & 1741 & $ < 13.024$ &  \nl
\tskip \tableline
\end{tabular}
\end{center}
\end{table}

\subsection{Q1759$+$75, $z$ = 2.625}

This system and the adopted $\N{HI} = 20.8 \pm 0.1$
are taken from an ongoing survey by I. Hook.
Figure~\ref{vpQ1759} presents the velocity plots and Table~\ref{Q1759-clm}
gives the measurements.
The QSO is very bright and was observed at FWHM $\approx 6 \mkms$
resolution.  We expect all of the measured abundances are
very accurate with the exception of $\N{Zn^+}$ where the ZnII~2026
profile is blended with sky lines.  
Although the strongest feature appears unblended,
we have chosen not to include Zn in the abundance analysis.

\begin{table}[ht] \footnotesize
\begin{center}
\caption{ \label{Q1759-clm}}
{\sc Ionic Column Densities: Q1759$+$75, $z$ = 2.625} 
\begin{tabular}{lcccc}
\tableline
\tableline \tskip
Ion & $\lambda$ & AODM & $N_{adopt}$ & [X/H] \\
\tableline \tskip
 HI & 1215 & $20.800 \pm 0.100$ &  \nl
 CIV & 1548 & $14.636 \pm 0.019$ &  \nl
 & 1550 & $14.647 \pm 0.005$ &  \nl
 AlIII & 1854 & $13.623 \pm 0.004$ &  \nl
 SiII & 1526 & $ > 15.014$ & $15.532 \pm 0.008$ & $-0.818 \pm 0.100$ \nl
 & 1808 & $15.532 \pm 0.008$ &  \nl
 CrII & 2066 & $13.211 \pm 0.062$ & $13.211 \pm 0.062$ & $-1.269 \pm 0.117$ \nl
 FeII & 1608 & $ > 14.980$ & $15.076 \pm 0.042$ & $-1.234 \pm 0.108$ \nl
 & 1611 & $15.076 \pm 0.042$ &  \nl
 NiII & 1454 & $13.589 \pm 0.039$ & $13.565 \pm 0.011$ & $-1.485 \pm 0.101$ \nl
 & 1709 & $13.615 \pm 0.020$ &  \nl
 & 1741 & $13.582 \pm 0.017$ &  \nl
 & 1751 & $13.506 \pm 0.021$ &  \nl
 ZnII & 2026 & $ > 11.650$ & $> 11.650$ & $> -1.800$ \nl
\tskip \tableline
\end{tabular}
\end{center}
\end{table}

\subsection{Q2230$+$02, $z$ = 1.864}

Figure~\ref{vpQ2230} presents the velocity plots for the damped \lya
system toward Q2230$+$02 and Table~\ref{Q2230-clm} lists the
measured ionic column densities.
We adopt $\log \N{HI} = 20.85 \pm 0.084$ (\cite{ptt94}) for this
LBQS damped system.   We have wavelength coverage of 28 metal-line
transitions and have determined accurate column densities for the majority
of them.  This system is notable for a $4 \sigma$ detection of
$\N{Ti^+}$. 
There are two TiII transitions at 1910~\rAA 
with similar oscillator strengths and the profiles cannot
be disentangled.
Following the analysis presented in Prochaska \& Wolfe (1997a), 
we integrate $N_a$
over the entire velocity region associated with the TiII
transitions and use a reduced oscillator strength, 
$f^* \equiv (f_1^2 + f_2^2) / (f_1 + f_2) = 0.0862$, 
to calculate $\N{Ti^+}$.  There is another difficulty
in determining the Ti$^+$ column density for this system.
An emission line feature lies
just blueward of the TiII profiles and causes problems
in determining the continuum level.
This leads to a systematic error estimated at 0.1 dex.

The profile for the low-ion transitions is comprised of three primary
features whose relative optical depth apparently varies from transition
to transition. For instance, the central feature is the strongest 
in the SiII~1808, CrII~2062, CrII~2066, and ZnII~2026 profiles and 
the weakest in the FeII~2260 and CrII~2056 transitions.  
While differences in the level of 
dust depletion could account for some of the
variations, it would be impossible to explain the differences for a
single ion
(e.g.\ the CrII transitions).  The most likely explanation
is that this central feature is a very narrow component -- not
fully resolved at our resolution -- whose relative optical depth 
varies significantly with the strength of the transition.
Finally, this system is notable for the presence of a nearby metal-line
system at $\approx -550 \mkms$ which shows multiple metal transitions
(Figure~\ref{vpQ2230B}).   
The subsystem exhibits ionic column densities typical of a damped
\lya system and its Al$^+$/Al$^{++} = 2.6$ ratio indicates the system is
nearly neutral.  Therefore, it is almost certainly contributing to the
measured $\N{HI}$ for this system which would mean we are underestimating
the metallicity of the system at $z = 1.864$.  Because the nearby
system exhibits ionic column densities at $< 10 \%$ of the 
damped system, we have chosen to ignore its effects in our 
abundance analysis.

\begin{table}[ht] \footnotesize
\begin{center}
\caption{ \label{Q2230-clm}}
{\sc Ionic Column Densities: Q2230$+$02, $z$ = 1.864} 
\begin{tabular}{lcccc}
\tableline
\tableline \tskip
Ion & $\lambda$ & AODM & $N_{adopt}$ & [X/H] \\
\tableline \tskip
 HI & 1215 & $20.850 \pm 0.084$ &  \nl
 CIV & 1548 & $ > 14.848$ &  \nl
 & 1550 & $14.808 \pm 0.007$ &  \nl
 AlII & 1670 & $ > 14.020$ &  \nl
 AlIII & 1854 & $13.590 \pm 0.006$ &  \nl
 & 1862 & $13.596 \pm 0.009$ &  \nl
 SiII & 1526 & $ > 15.233$ & $15.656 \pm 0.010$ & $-0.744 \pm 0.085$ \nl
 & 1808 & $15.656 \pm 0.010$ &  \nl
 SiIV & 1393 & $ > 14.267$ &  \nl
 & 1402 & $ > 14.354$ &  \nl
 TiII & 1910 & $12.985 \pm 0.099$ & $12.985 \pm 0.099$ & $-0.795 \pm 0.130$ \nl
 CrII & 2056 & $13.361 \pm 0.036$ & $13.403 \pm 0.027$ & $-1.127 \pm 0.088$ \nl
 & 2066 & $13.483 \pm 0.041$ &  \nl
 FeII & 1608 & $ > 15.115$ & $15.188 \pm 0.016$ & $-1.172 \pm 0.086$ \nl
 & 1611 & $15.273 \pm 0.084$ &  \nl
 & 2249 & $15.120 \pm 0.036$ &  \nl
 & 2260 & $15.210 \pm 0.019$ &  \nl
 & 2344 & $ > 14.988$ &  \nl
 & 2374 & $ > 15.183$ &  \nl
 & 2382 & $ > 14.723$ &  \nl
 NiII & 1370 & $13.883 \pm 0.052$ & $13.812 \pm 0.011$ & $-1.288 \pm 0.085$ \nl
 & 1709 & $13.860 \pm 0.014$ &  \nl
 & 1741 & $13.838 \pm 0.023$ &  \nl
 & 1751 & $13.687 \pm 0.028$ &  \nl
 ZnII & 2026 & $12.800 \pm 0.028$ & $12.800 \pm 0.028$ & $-0.700 \pm 0.088$ \nl
\tskip \tableline
\end{tabular}
\end{center}
\end{table}

\subsection{Q2231$-$00, $z$ = 2.066}

This LBQS system is another damped \lya system exhibiting significant
$(5 \sigma)$ TiII~1910
absorption.    As in the damped \lya system toward
Q2230$+$02, the TiII profiles overlap,
hence we use the technique outlined above to determine $\N{Ti^+}$.
Throughout the analysis we adopt 
$\log \N{HI} = 20.56 \pm 0.1$ (\cite{ptt94}).
The velocity plots are given in Figure~\ref{vpQ2231} and the
column densities are presented in Table~\ref{Q2231-clm}.
This is the other system in our data sample which was previously
observed by Lu et al.\ (1996). 
We find our measurements match theirs in nearly every case, 
although their analysis did not include Ti$^+$.

\begin{table}[ht] \footnotesize
\begin{center}
\caption{ \label{Q2231-clm}}
{\sc Ionic Column Densities: Q2231$-$00, $z$ = 2.066} 
\begin{tabular}{lcccc}
\tableline
\tableline \tskip
Ion & $\lambda$ & AODM & $N_{adopt}$ & [X/H] \\
\tableline \tskip
 HI & 1215 & $20.560 \pm 0.100$ &  \nl
 CII & 1335 & $13.580 \pm 0.040$ &  \nl
 CIV & 1548 & $14.195 \pm 0.005$ &  \nl
 OI & 1302 & $ > 15.543$ & $< 17.748$ & $<  0.258$ \nl
 & 1355 & $ < 17.748$ &  \nl
 AlIII & 1854 & $13.172 \pm 0.010$ &  \nl
 & 1862 & $13.110 \pm 0.023$ &  \nl
 SiII & 1304 & $ > 15.030$ & $15.247 \pm 0.019$ & $-0.863 \pm 0.102$ \nl
 & 1526 & $ > 15.014$ &  \nl
 & 1808 & $15.247 \pm 0.019$ &  \nl
 SiIV & 1393 & $13.704 \pm 0.010$ &  \nl
 & 1402 & $13.698 \pm 0.012$ &  \nl
 TiII & 1910 & $12.848 \pm 0.071$ & $12.848 \pm 0.071$ & $-0.642 \pm 0.123$ \nl
 CrII & 2062 & $13.182 \pm 0.040$ & $13.165 \pm 0.034$ & $-1.075 \pm 0.106$ \nl
 & 2066 & $13.130 \pm 0.063$ &  \nl
 FeII & 1608 & $14.750 \pm 0.009$ & $14.750 \pm 0.009$ & $-1.320 \pm 0.100$ \nl
 & 1611 & $14.783 \pm 0.065$ &  \nl
 NiII & 1370 & $12.880 \pm 0.091$ & $13.306 \pm 0.032$ & $-1.504 \pm 0.105$ \nl
 & 1741 & $13.381 \pm 0.033$ &  \nl
 & 1751 & $13.100 \pm 0.091$ &  \nl
 ZnII & 2026 & $12.463 \pm 0.023$ & $12.463 \pm 0.023$ & $-0.747 \pm 0.103$ \nl
\tskip \tableline
\end{tabular}
\end{center}
\end{table}

\subsection{Q2348$-$14, $z$ = 2.279}

This very metal-poor system was first discussed by Pettini et al.\
(1994) and has 
been previously observed at high resolution by Pettini et al.\ (1995).
In Figure~\ref{Q2348-lya} we plot the \lya transition for this
system.  Overplotted are Voigt profiles centered at $z = 2.2794$
with $\log \N{HI} = 20.56 \pm 0.075$ corresponding to
the measurements made by Pettini et al.\ (1994).
While the profiles provide a reasonable fit to the left wing
of the damped profile, the fit for $v > 750 \mkms$ is clearly
inconsistent for all of the assumed $\N{HI}$ values.  It is very
difficult, however, to continuum fit an order of HIRES data
which includes a damped \lya profile; it is easier
to fit intermediate resolution data. 
Therefore, we adopt $\log \N{HI} = 20.56 \pm 0.075$
from the Pettini et al.\ analysis, but note in passing that this
may be an overestimate of the true $\N{HI}$ value.
Comparing our derived chemical 
abundances with the work of Pettini et al.\ (1995)
we find reasonable agreement and note we 
have improved on their limits in a few cases (e.g.\ N and S).

The system is exceptional for a number of reasons.
First, it is one of the few systems where we have a measurement
of $\N{S^+}$ based on the SII 1259 transition.  As discussed for the
damped system toward Q0347$-$38, S/Fe is an
excellent diagnostic of dust depletion.
Here we find [S/Fe] = 0.17~dex which argues strongly
against dust depletion
{\it if} the system has an underlying Type II SN pattern.
Second, the system exhibits two distinct low-ion features, one 
at $v \approx -100 \mkms$ (feature 1; this feature was not resolved
in previous observations)
and the more dominant at $v \approx 0 \mkms$ (feature 2).
Feature 1 is present in all of the SiII transitions and is the strongest
feature in the AlIII, CIV and SiIV profiles.
Comparing $\N{Si^+}$ for the two features in the SiII 1526 profile,
we find $\N{Si^+}_1 / \N{Si^+}_2 = 0.18$.  
At the same time, however, 
the feature is entirely absent in the OI~1302 and FeII~1608 profiles: 
$\N{O^0}_1 / \N{O^0}_2 < 0.027$ and 
$\N{Fe^+}_1 / \N{Fe^+}_2 < 0.07$.
While dust could possibly explain the absence of Fe where Si is
present because Si is only lightly depleted in the ISM,
it cannot account for the lack of OI absorption.
The system corresponding to feature~1 must be significantly
ionized such that the dominant states of O, Fe, and Si are
higher ions.  
We have performed CLOUDY (\cite{fer91}) calculations which demonstrate
values of [Si$^+$/O$^0$]~$\approx 1.2$~dex
and [Si$^+$/Fe$^+$] $\approx 0.7$~dex are possible in a Lyman limit
system provided the ionization parameter is significantly high. Therefore, we 
argue feature 2 marks the damped \lya system while feature
1 may be a significantly ionized satellite or halo cloud.
The system is also notable for providing a meaningful estimate of N/O.
Finally, perhaps the most remarkable characteristic of this damped system
is the velocity width of the 
CIV profiles, $\delv > 600 \mkms$, particularly
in light of the very narrow $\delv$ exhibited by the low-ion profiles.
This kinematic observation poses
a difficult challenge to any physical model introduced
to explain the damped \lya systems.

\begin{table}[ht] \footnotesize
\begin{center}
\caption{ \label{Q2348B-clm}}
{\sc Ionic Column Densities: Q2348$-$14, $z$ = 2.279} 
\begin{tabular}{lcccc}
\tableline
\tableline \tskip
Ion & $\lambda$ & AODM & $N_{adopt}$ & [X/H] \\
\tableline \tskip
 HI & 1215 & $20.560 \pm 0.075$ &  \nl
 CII & 1334 & $ > 14.610$ &  \nl
 & 1335 & $13.207 \pm 0.070$ &  \nl
 NI & 1200 & $ < 13.223$ & $< 13.223$ & $< -3.387$ \nl
 OI & 1302 & $ > 14.798$ & $> 14.798$ & $> -2.692$ \nl
 AlII & 1670 & $12.654 \pm 0.006$ & $12.654 \pm 0.006$ & $-2.386 \pm 0.075$ \nl
 AlIII & 1854 & $12.726 \pm 0.012$ &  \nl
 & 1862 & $12.613 \pm 0.027$ &  \nl
 SiII & 1190 & $ > 14.207$ & $14.201 \pm 0.010$ & $-1.909 \pm 0.076$ \nl
 & 1193 & $ > 13.924$ &  \nl
 & 1260 & $ > 13.744$ &  \nl
 & 1304 & $14.227 \pm 0.021$ &  \nl
 & 1526 & $14.201 \pm 0.011$ &  \nl
 & 1808 & $14.089 \pm 0.059$ &  \nl
 SiIV & 1393 & $ > 14.063$ &  \nl
 SII & 1259 & $13.725 \pm 0.119$ & $13.725 \pm 0.119$ & $-2.105 \pm 0.140$ \nl
 FeII & 1608 & $13.792 \pm 0.016$ & $13.792 \pm 0.016$ & $-2.278 \pm 0.077$ \nl
\tskip \tableline
\end{tabular}
\end{center}
\end{table}

\subsection{Q2359$-$02, $z$=2.095 \& $z$=2.154}

This faint QSO exhibits two intervening damped \lya systems.
Both are members of the LBQS statistical survey.
The velocity plots and column densities for the $z=2.095$ system
are presented in Figure~\ref{vpQ2359A} and Table~\ref{Q2359A-clm}
and those for the $z=2.154$ are given by Figure~\ref{vpQ2359B}
and Table~\ref{Q2359B-clm}.
Both are part of the LBQS statistical sample and we have taken
the HI column densities from Wolfe et al.\ (1995):
$\log \N{HI} = 20.7 \pm 0.1$ for the $z=2.095$ system and
$\log \N{HI} = 20.3 \pm 0.1$ for the $z=2.154$ system.
The SNR is relatively low for most of the spectrum and therefore
abundances established on the weakest transitions are suspect,
particularly those for Zn.  We intend to make further observations of
these two systems to improve the accuracy of our abundance measurements.

\begin{table}[ht] \footnotesize
\begin{center}
\caption{ \label{Q2359A-clm}}
{\sc Ionic Column Densities: Q2359$-$02, $z$ = 2.095} 
\begin{tabular}{lcccc}
\tableline
\tableline \tskip
Ion & $\lambda$ & AODM & $N_{adopt}$ & [X/H] \\
\tableline \tskip
 HI & 1215 & $20.700 \pm 0.100$ &  \nl
 CII & 1334 & $ > 15.146$ &  \nl
 CIV & 1548 & $ > 14.639$ &  \nl
 & 1550 & $ > 14.800$ &  \nl
 AlII & 1670 & $13.662 \pm 0.034$ & $13.662 \pm 0.034$ & $-1.518 \pm 0.106$ \nl
 AlIII & 1854 & $13.383 \pm 0.010$ &  \nl
 & 1862 & $13.396 \pm 0.016$ &  \nl
 SiII & 1526 & $ > 15.045$ & $15.408 \pm 0.021$ & $-0.842 \pm 0.102$ \nl
 & 1808 & $15.408 \pm 0.021$ &  \nl
 SiIV & 1393 & $ > 14.072$ &  \nl
 & 1402 & $ > 14.517$ &  \nl
 CrII & 2056 & $12.748 \pm 0.091$ & $12.748 \pm 0.091$ & $-1.632 \pm 0.135$ \nl
 & 2062 & $13.131 \pm 0.057$ &  \nl
 & 2066 & $ < 12.897$ &  \nl
 FeII & 1608 & $14.507 \pm 0.025$ & $14.507 \pm 0.025$ & $-1.703 \pm 0.103$ \nl
 & 1611 & $ < 15.077$ &  \nl
 NiII & 1703 & $ < 13.693$ & $13.142 \pm 0.054$ & $-1.808 \pm 0.114$ \nl
 & 1709 & $13.152 \pm 0.127$ &  \nl
 & 1741 & $13.185 \pm 0.072$ &  \nl
 & 1751 & $13.071 \pm 0.107$ &  \nl
 ZnII & 2026 & $12.595 \pm 0.029$ & $12.595 \pm 0.029$ & $-0.755 \pm 0.104$ \nl
 & 2062 & $12.486 \pm 0.075$ &  \nl
\tskip \tableline
\end{tabular}
\end{center}
\end{table}

\begin{table}[ht] \footnotesize
\begin{center}
\caption{ \label{Q2359B-clm}}
{\sc Ionic Column Densities: Q2359$-$02, $z$ = 2.154} 
\begin{tabular}{lcccc}
\tableline
\tableline \tskip
Ion & $\lambda$ & AODM & $N_{adopt}$ & [X/H] \\
\tableline \tskip
 HI & 1215 & $20.300 \pm 0.100$ &  \nl
 CII & 1334 & $ > 14.987$ &  \nl
 CIV & 1548 & $ > 14.880$ &  \nl
 & 1550 & $ > 14.966$ &  \nl
 AlII & 1670 & $13.165 \pm 0.016$ & $13.165 \pm 0.016$ & $-1.615 \pm 0.101$ \nl
 AlIII & 1862 & $12.960 \pm 0.021$ &  \nl
 SiII & 1526 & $14.316 \pm 0.015$ & $14.316 \pm 0.015$ & $-1.534 \pm 0.101$ \nl
 SiIV & 1393 & $ > 14.568$ &  \nl
 & 1402 & $ > 14.528$ &  \nl
 CrII & 2056 & $ < 12.549$ & $< 12.946$ & $< -1.034$ \nl
 & 2066 & $ < 12.946$ &  \nl
 FeII & 1608 & $13.895 \pm 0.032$ & $13.895 \pm 0.032$ & $-1.915 \pm 0.105$ \nl
 & 1611 & $ < 14.917$ &  \nl
 NiII & 1317 & $ < 13.152$ & $< 12.949$ & $< -1.601$ \nl
 & 1370 & $ < 13.248$ &  \nl
 & 1454 & $ < 13.063$ &  \nl
 & 1741 & $ < 12.949$ &  \nl
 & 1751 & $ < 13.200$ &  \nl
 ZnII & 2026 & $ < 11.901$ & $< 11.901$ & $< -1.049$ \nl
\tskip \tableline
\end{tabular}
\end{center}
\end{table}

\setcounter{figure}{21}
\begin{figure*}
\begin{center}
\includegraphics[height=8.5in, width=5.3in, bb = 55 48 557 744]{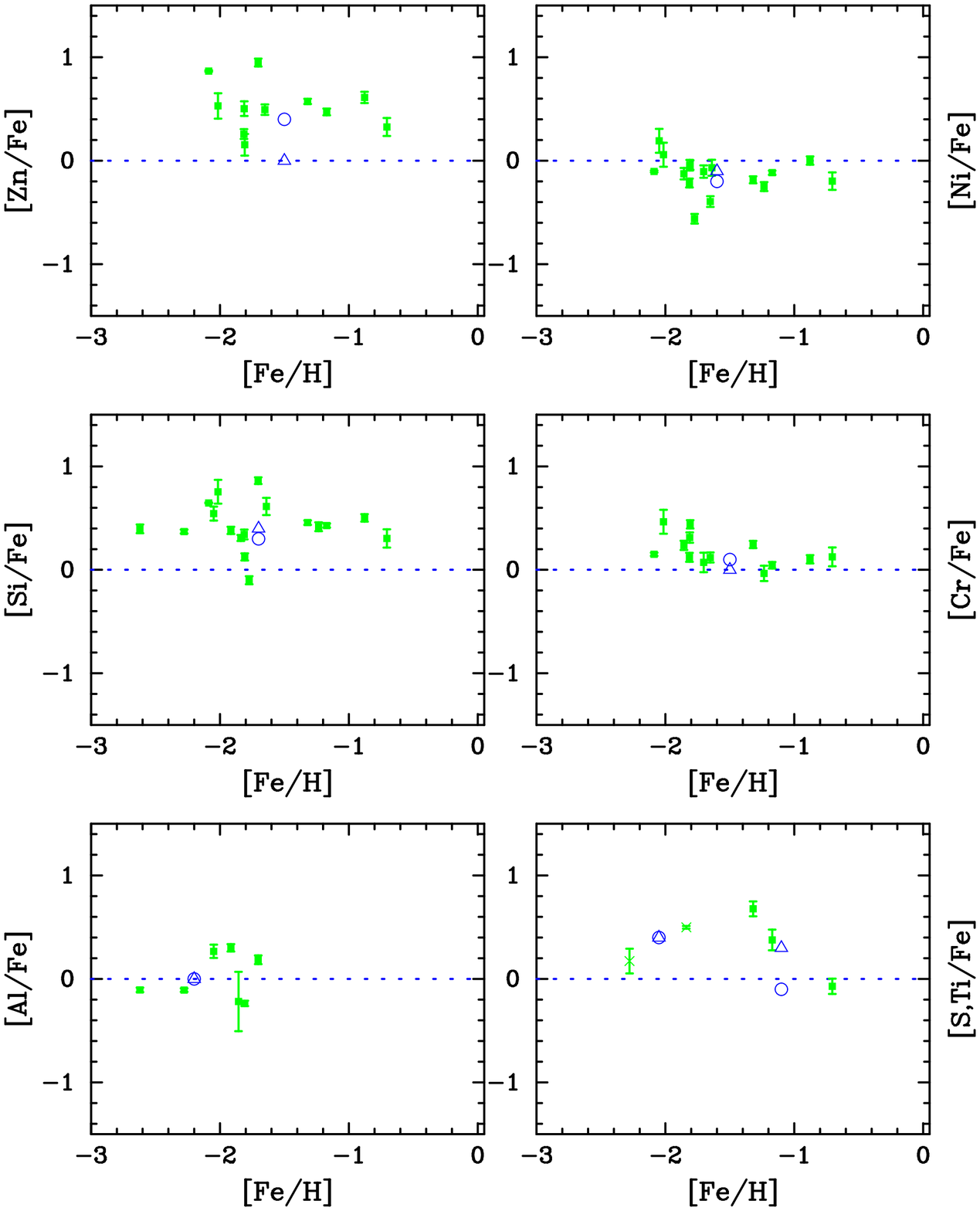}
\caption{ Abundance patterns of the most common elements from our full
damped \lya sample.  Following standard practice in stellar abundance
analysis we plot [X/Fe], the logarithmic abundance of element X to
Fe relative to solar.  We plot this quantity vs.\ [Fe/H], an indicator
of the metallicity of the system.  The triangles printed in each panel
represent the typical metal-poor halo value (\cite{mcwll97}).  
The circles indicate the observed
pattern for lightly depleted ISM gas (\cite{sav96}).  In the lower right
panel, the $x$'s identify [S/Fe] values and the squares mark
[Ti/Fe]. }
\label{abd-patt}
\end{center}
\end{figure*}

\begin{sidewaystable}
\begin{center}
\caption{ \label{abnd} }
\begin{tabular}{lccccccccc}
\tableline
\tableline \tskip
DLA & $z_{abs}$ & log[N(HI)] &  [Si/H] & [Al/H] & [S/H] & [Fe/H] & [Ni/H] & [Cr/H] & [Zn/H] \\
\tableline \tskip
 Q0000$-$26 & 3.390 & 21.410 & $ -1.874$ &  &  & $ -1.774$ & $ -2.335$ &  &  \\ 
 Q0019$-$01 & 3.439 & 20.900 & $ -1.027$ &  &  & $ -1.640$ & $ -1.708$ &  &  \\ 
 Q0100+13 & 2.309 & 21.400 & $  > -2.228$ &  &  & $ -1.814$ & $ -2.030$ & $ -1.693$ & $ -1.556$ \\ 
 Q0149+33 & 2.141 & 20.500 & $ -1.683$ & $ -2.044$ &  & $ -1.808$ & $ -1.850$ & $ -1.369$ & $ -1.654$ \\ 
 Q0201+36 & 2.463 & 20.380 & $ -0.376$ &  &  & $ -0.878$ & $ -0.877$ & $ -0.776$ & $ -0.266$ \\ 
 Q0347$-$38 & 3.025 & 20.800 & $ -1.530$ &  & $ -1.339$ & $ -1.838$ & $  < -2.373$ &  &  \\ 
 Q0458$-$02 & 2.040 & 21.650 & $  > -2.105$ &  &  & $ -1.652$ & $ -2.047$ & $ -1.533$ & $ -1.159$ \\ 
 Q0841+12A & 2.375 & 20.950 & $ -1.261$ & $  > -2.163$ &  & $ -2.015$ & $ -1.957$ & $ -1.551$ & $ -1.485$ \\ 
 Q0841+12B & 2.476 & 20.780 & $  > -1.869$ & $ -2.074$ &  & $ -1.856$ & $ -1.982$ & $ -1.620$ & $  < -1.652$ \\ 
 Q0951$-$04A & 3.857 & 20.600 & $ -1.505$ & $ -1.782$ &  & $ -2.048$ & $ -1.856$ &  &  \\ 
 Q0951$-$04B & 4.203 & 20.400 & $ -2.558$ &  &  & $  < -2.629$ & $  < -2.061$ &  &  \\ 
 Q1215+33 & 1.999 & 20.950 & $ -1.470$ &  &  & $ -1.812$ & $ -1.856$ & $ -1.500$ & $ -1.309$ \\ 
 Q1331+17 & 1.776 & 21.176 & $ -1.441$ &  &  & $ -2.088$ & $ -2.191$ & $ -1.937$ & $ -1.221$ \\ 
 Q1346$-$03 & 3.736 & 20.700 & $ -2.296$ & $ -2.634$ &  &  &  &  &  \\ 
 Q1759+75 & 2.625 & 20.800 & $ -0.818$ &  &  & $ -1.234$ & $ -1.485$ & $ -1.269$ & $  > -1.800$ \\ 
 Q2206$-$19A & 1.920 & 20.653 & $ -0.402$ &  &  & $ -0.705$ & $ -0.903$ & $ -0.580$ & $ -0.379$ \\ 
 Q2206$-$19B & 2.076 & 20.431 & $ -2.225$ & $ -2.727$ &  & $ -2.621$ &  &  &  \\ 
 Q2230+02 & 1.864 & 20.850 & $ -0.744$ &  &  & $ -1.172$ & $ -1.288$ & $ -1.127$ & $ -0.700$ \\ 
 Q2231$-$00 & 2.066 & 20.560 & $ -0.863$ &  &  & $ -1.320$ & $ -1.504$ & $ -1.075$ & $ -0.747$ \\ 
 Q2348$-$14 & 2.279 & 20.560 & $ -1.909$ & $ -2.386$ & $ -2.105$ & $ -2.278$ &  &  &  \\ 
 Q2359$-$02A & 2.095 & 20.700 & $ -0.842$ & $ -1.518$ &  & $ -1.703$ & $ -1.808$ & $ -1.632$ & $ -0.755$ \\ 
 Q2359$-$02B & 2.154 & 20.300 & $ -1.534$ & $ -1.615$ &  & $ -1.915$ & $  < -1.601$ & $  < -1.034$ & $  < -1.049$ \\ 
\tskip \tableline
\end{tabular}
\end{center}
\end{sidewaystable}

\section{ABUNDANCE PATTERNS}
\label{sec-abnd}

Table~\ref{abnd} lists the relative logarithmic
abundances for the damped \lya systems, 
[X/H] $\equiv \log[\N{X}/\N{H}] - \log[\N{X}/\N{H}]_\odot$.
For each system we adopt the value for $\N{HI}$ indicated
in the previous section, and calculate $\N{X}$ by performing
a weighted mean of all direct measurements, i.e.\ no limits or blends. 
Included in Table~\ref{abnd} are the abundance measurements
for the damped \lya systems toward Q0201$+$36 and Q2206$-$19.
In a few cases the values differ from the published values
in light of new oscillator strengths.
It is well-established both
observationally and theoretically that the damped 
\lya systems are primarily neutral (\cite{vgs94,pro96}).
Therefore we perform no ionization corrections in 
determining the abundances from the ionic column densities 
of the low-ions.

Figure~\ref{abd-patt} presents the abundance patterns for the most
common elements in our full sample.  We have chosen not to include
error bars for presentation purposes;  the derived errors are 
$< 0.1$ dex with only a few exceptions. 
Following Lu et al.\ (1996), 
along the x-axis of each panel we plot metallicity --
here expressed with [Fe/H] -- in part because this is the primary metallicity
indicator in local stellar populations and in part because we have
Fe abundance measurements for a greater number of systems.
Examining the upper-left panel of Figure~\ref{abd-patt}, 
we note a systematic overabundance of Zn/Fe
which suggests Fe may be depleted onto dust grains.  Therefore consider the
[Fe/H] values to be lower limits to the true metallicity.
In one case where (Q0841+12A)
we have no reliable Fe abundance measurement and have taken 
[Fe/H] = [Cr/H] on the grounds that [Cr/H] $\approx$ [Fe/H]
in the majority of damped \lya systems.

In the following, we
discuss the evidence for dust depletion and Type II supernovae
enrichment in light of the observed damped \lya abundance patterns.
The former is determined by comparing against abundance patterns in 
depleted ISM clouds (\cite{sav96})
while the latter is assessed by comparing against
the abundance patterns of metal-poor halo stars presumed to exhibit
nucleosynthetic patterns typical of Type II supernovae (\cite{mcwll97}).
First, consider the top two panels of Figure~\ref{abd-patt}
which lend support to the
presence of significant dust depletion in the damped \lya systems.
As described throughout the paper, the overabundance of Zn 
to Fe relative to solar abundances is suggestive of dust depletion
both because (i) Zn is largely undepleted in dusty regions within
the ISM whereas Fe is heavily depleted
and (ii) [Zn/Fe] $\approx 0$~dex for stars of all metallicity
observed within the Galaxy (\cite{sne91}).  Similarly, the
Ni/Fe ratio is significantly lower 
than metal-poor halo stars, which is consistent with Ni 
being more heavily depleted than Fe in
depleted regions of the ISM.  
Lu et al.\ (1996) have argued the underabundance
is primarily due to an error in the oscillator strengths for the
NiII transitions.  While a recent analysis by Zsarg$\rm \acute o$
\& Federman (1998) indicates the NiII $f$-values are poorly determined,
it is not clear if this can entirely account for the discrepancy between
the damped \lya observations and the metal-poor halo star abundances.
For our analysis, we have adopted the updated
$f$-values from Zsarg$\rm \acute o$ \& Federman (1998) -- which does
include a decrease in $f_{1714}$ by a factor of 1.34 -- yet a significant
underabundance of [Ni/Fe] is still apparent.  Therefore it is unclear
if errors in the oscillator strengths can fully resolve the discrepancy
between the observed Ni/Fe pattern and that predicted for Type II SN
yields.

In contrast to the top two panels, the middle panels and the Al/Fe
abundance pattern are generally consistent
with both dust depletion and Type II SN enrichment.
As emphasized by Lu et al.\ (1996), the overabundance\footnote{The one data
point with [Si/Fe] $< 0$ is from Q0000$-$26 where both the Si$^+$ and Fe$^+$
column densities are insecure.} of Si/Fe relative
to solar is very suggestive of Type II supernovae enrichment (\cite{mcwll97}).
In the case of Si/Fe, the Si overabundance is explained as the result
of the overproduction of Si -- an $\alpha$-element relative to Fe in
Type II supernovae.
Similarly, the Cr/Fe and Al/Fe patterns are consistent with those
observed for the metal-poor halo stars.
Contrary to the Lu et al.\ observations, however, we observe
an overabundance of Cr/Fe at very low metallicity
(for [Fe/H]~$<-1.5$, $<$[Cr/Fe]$> = +0.21$).  
This result is most likely due to the fact that we are biased
to high [Cr/Fe] values at low [Fe/H] because low [Cr/Fe] values
would imply Cr$^+$ column densities below our detection limit.
In one system (Q0149+33)
we observe [Cr/Zn]~$> 0$~dex indicating it is essentially undepleted by
dust grains.  Furthermore,  we measure an overabundance of Si relative
to Fe in this system ([Si/Fe] = $+0.12$~dex) which is an indication
of a Type II $\alpha$-enhancement although at a somewhat smaller level
than most metal-poor halo stars. Lastly, while
the [Al/Fe] measurements are broadly consistent with the abundances
observed in metal-poor halo stars, there may be a contradiction at
very low [Fe/H].
For halo stars with [Fe/H] $< -2$~dex, [Al/Fe] $< -0.4$~dex; \cite{mcwll97})
yet if anything the damped \lya systems exhibit [Al/Fe] $> 0$~dex at
this metallicity.  This result may ultimately pose a serious challenge to the
interpretation of Type II SN nucleosynthetic patterns.

\begin{figure}[ht]
\begin{center}
\includegraphics[height=3.5in, width=3.0in, angle=-90, bb = 65 58 557 734]{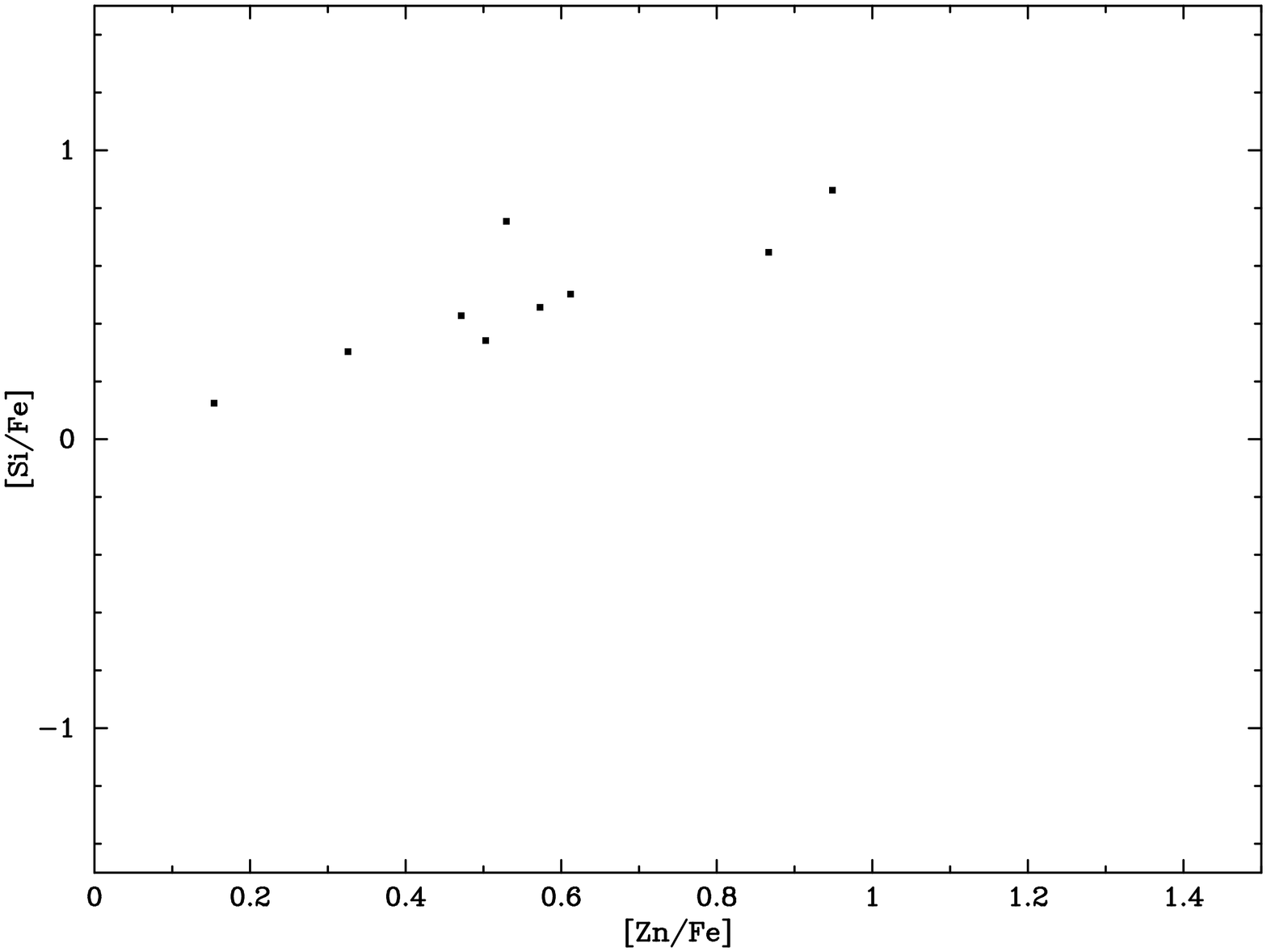}
\caption{ [Si/Fe], [Zn/Fe] pairs for the 9 damped \lya systems 
from the full sample exhibiting Si, Zn, and Fe.  The positive correlation
(Pearson's correlation coeffecient $r = 0.86$ with the null hypothesis 
a 0.003 probability) is suggestie of dust depletion, but could possibly
have a nucleosynthetic origin (see text).}
\label{dustplt}
\end{center}
\end{figure}

While the abundances for Cr, Al and Si vs.\ Fe
resemble those for the metal-poor
halo stars, the patterns also tend to match the dust depletion patterns
of lightly depleted regions within the ISM (\cite{sav96}). 
In these regions, Si is overabundant relative to Fe,
[Cr/Fe]~$\approx 0.1$~dex, and 
recent measurements of the Al to Fe ratio towards three OB stars
(\cite{howk98}) suggest [Al/Fe]~$\gtrsim 0$~dex.
If the overabundance of Si/Fe relative to solar is indicative of dust
depletion, then one might expect a correlation between [Si/Fe] and 
[Zn/Fe] with the most heavily depleted regions showing the largest
Si/Fe and Zn/Fe ratios.  
A plot of [Si/Fe] vs.\
[Zn/Fe] for all the systems with accurate abundances for the three elements
(Figure~\ref{dustplt})  reveals a positive correlation
(the Pearson coeffecient is 0.86 in log-space with a null hypothesis
probability of 0.003), consistent
with that expected for dust depletion. 
However, if Zn is produced in the neutrino-driven winds of Type II
SN (\cite{hff96}), one may also expect a correlation between the 
abundance of Si and Zn relative to Fe.

Now consider the observations of Ti/Fe (solid squares in 
the lower right hand panel) which pose a strong
argument for Type II SN enrichment.
As emphasized in Prochaska \& Wolfe (1997a),
Ti is more heavily depleted than Fe in dusty regions within the ISM
(\cite{lpp95}),
yet we find [Ti/Fe]~$\geq 0$ in every damped \lya systems where Ti
is observed.
As Ti is an $\alpha$-element, this argues strongly for the
Type II SN interpretation.
Lu et al.\ (1996) have made similar arguments 
for the observed underabundance of Mn/Fe in the damped systems.
Because Mn is less depleted than Fe in dusty regions of the ISM,
the Mn/Fe underabundance cannot be explained by dust depletion.
On the other hand one observes
[Mn/Fe]$< 0$~dex for the metal-poor halo stars (\cite{mcwll97}).
Furthermore, the [Mn/Fe] values show a similar trend with [Fe/H] (albeit
in terms of an underabundance) to that of the $\alpha$-elements. 
It is possible this trend indicates
a metallicity dependent yield for Mn, but the plateau
in [Mn/Fe] values at [Fe/H] $\approx -1$~to~$-2.5$ is better understood
if Mn is overproduced relative to Fe by Type Ia supernovae 
(\cite{nak98}).  If the latter explanation is correct, then
the low [Mn/Fe] values are significant evidence for Type II SN 
enrichment within the damped \lya systems.
At the very least, we wish to stress the damped [Mn/Fe] observations 
require that the underlying nucleosynthetic pattern
does not simply match solar abundances.

For the elements considered thus far, the abundance patterns are 
broadly consistent with a combination of Type II SN enrichment
and an 'ISM-like' dust depletion pattern.  This is not the case for
Sulphur.
In the two cases from our full sample where
we have accurate measurements for S/Fe we find:
(i) [S/Fe] = $0.50 \pm 0.02$ for the damped system at $z=3.025$ toward
Q0347$-$38 and (ii) [S/Fe] = $0.17 \pm 0.1$ for the damped system toward
Q2348$-$14.  Similar to Silicon, Sulfur is an $\alpha$-element and is 
observed to be overabundant relative to Fe in metal-poor halo stars by
[Si/Fe]$_{II} \approx 0.3 - 0.5$ dex.  
Like Zinc, Sulfur is undepleted in the ISM.
Therefore, interpreting the positive [Zn/Fe] values as the result of dust
depletion, one would expect typical values for 
[S/Fe]$_{dust}$~$> 0.3$~dex on the basis of depletion alone.
Given all of the damped systems -- 
including those from Lu et al.\ (1996) --
exhibit [S/Fe]~$\leq 0.5$~dex, the S abundance
pattern is inconsistent with a combination of 
Type II SN enrichment and dust depletion because this
would require [S/Fe]$_{obs}$ = [S/Fe]$_{II}$ + [S/Fe]$_{dust} > 0.6$~dex 
in every case.  
While this point has been discussed
previously, it needs to be emphasized.  If dust is playing the primary
role in the observed abundance patterns of the damped \lya systems,
then the [S/Fe] measurements require one of two conclusions:
(1) the damped \lya systems were {\it not} primarily enriched by
Type II SN or (2) all of the systems where S has been measured are
atypical in that they are the few which are undepleted.
The first conclusion is at odds with most theories of galactic chemical
evolution and is inconsistent with the observations of the Milky Way.
To adopt point (1), one would have to argue the chemical
history of the damped systems is very different from that of the Milky Way.
Point (2) is a possibility for Q2348$-$14, but the Ni/Fe ratio
observed for Q0347$-$38 ([Ni/Fe]~$< -0.5$) would indicate this system is
significantly depleted. At present, then,
any attempt to match the abundance patterns of the damped \lya systems
with a combination of Type II SN enrichment and ISM-like dust 
depletion must fail the S observations.

Synthesizing our results with those from previous studies,
we contend the abundance patterns of the damped
\lya systems lack any convincing single interpretation.  
On the face of it this may not be surprising, as one would
expect some differences in their chemical evolution.  While this
would explain variations of a particular X/Fe ratio, this
is unlikely to account for any of the inconsistencies discussed thus far.
While the majority of the 
patterns are in excellent agreement with the Type II SN
enriched halo star abundance
patterns, Zn/Fe and Ni/Fe are clearly inconsistent and are
very suggestive of dust depletion, albeit at considerably lower levels than
that observed in dusty ISM clouds.  
An 'ISM-like' dust depletion pattern on top of solar abundances accounts
for a majority of the observations but fails for Mn/Fe and Ti/Fe.
Attempts to match the
observed abundance patterns with a combination of dust depletion and
Type II supernovae enrichment have been largely unsuccessful
(\cite{lu96b,kulk97,vld98}).  
This failure is accentuated by our measurements of [S/Fe] which 
are inconsistent with a synthesis of dust depletion and Type~II SN
abundance patterns.  Of course to eliminate either effect would
have profound consequences.
If the damped \lya systems do not exhibit Type II SN abundances they
have a very different chemical evolution history than the Milky Way,
in that they do not match the stellar abundance patterns for
[Fe/H] $< -1$~dex.
This would beg the questions: Is the Milky Way unique? Or do
the damped \lya systems somehow not include the progenitors of present-day
spiral galaxies?  Also, why would the damped systems exhibit relative
solar abundances for all elements except Mn and Ti?
On the other hand,
if there is no dust depletion at play, then is the Zn/Fe ratio
observed in the Milky Way a special case?  
Also, are the [Al/Fe] observation at low metallicity consistent with
the Type II SN interpretation?

At the heart of these questions lies the physical nature of the
damped \lya systems.  If dust depletion is playing a principal role,
then perhaps the damped systems are tracing gas-rich galaxies not
unlike the Magellanic Clouds (\cite{wlty97}), whereas an underlying
Type II SN pattern is more suggestive of the progenitors of massive
spiral galaxies.
What steps can be taken to resolve these issues?  First, the 
overabudance of Ti/Fe must be confirmed.  A more accurate
measurement of the TiII~1910 $f$ values would be particularly
useful.  This could be achieved by performing observations of a system
showing both the TiII~1910 and TiII~3073 transitions.
Unfortunately, at present 
this requires high S/N, high resolution observations
at wavelengths exceeding 9000~\rAA (i.e.\ $z \approx 2$).  
One could much more easily
measure [Ti/Fe] in a few low $z$ systems via the TiII~3073 transition,
but the majority of these systems exhibit [Fe/H]~$> -1$~dex 
(\cite{ptt98}) and therefore
may not offer a fair comparison with the metal-poor halo stars.  It is 
interesting to note, however, that the presumed damped system
at $z =0.75$ towards Q2206$-$19 has [Ti/Fe]$= 0.27$~dex 
indicates an underlying Type II SN abundance pattern (\cite{pro97a}). 
Second, one can look to the relative abundances of the $\alpha$-elements
S, Si, O and Ti to investigate consistency with the metal-poor halo
star patterns.  Thus far the few data points we have are consistent
with the Type II SN interpretation  ([S/Si]$\approx$ 0.2 dex and
[Si/Ti] $\approx 0.1$~dex).  Third, the conclusions we have drawn
from the S/Fe ratio are based on very few damped \lya systems.
Given the importance of this particular ratio, further measurements
of Sulphur would be particularly enlightening.  Lastly, a 
more detailed abundance analysis of the lowest metallicity systems
would allow one to investigate the chemical evolution of the damped
systems.  Assuming the extremely metal-poor systems
are the least depleted, they should provide the most accurate indication
of the underlying nucleosynthetic pattern.

\begin{figure}[ht]
\begin{center}
\includegraphics[height=3.5in, width=3.0in, angle=-90, bb = 65 58 557 734]{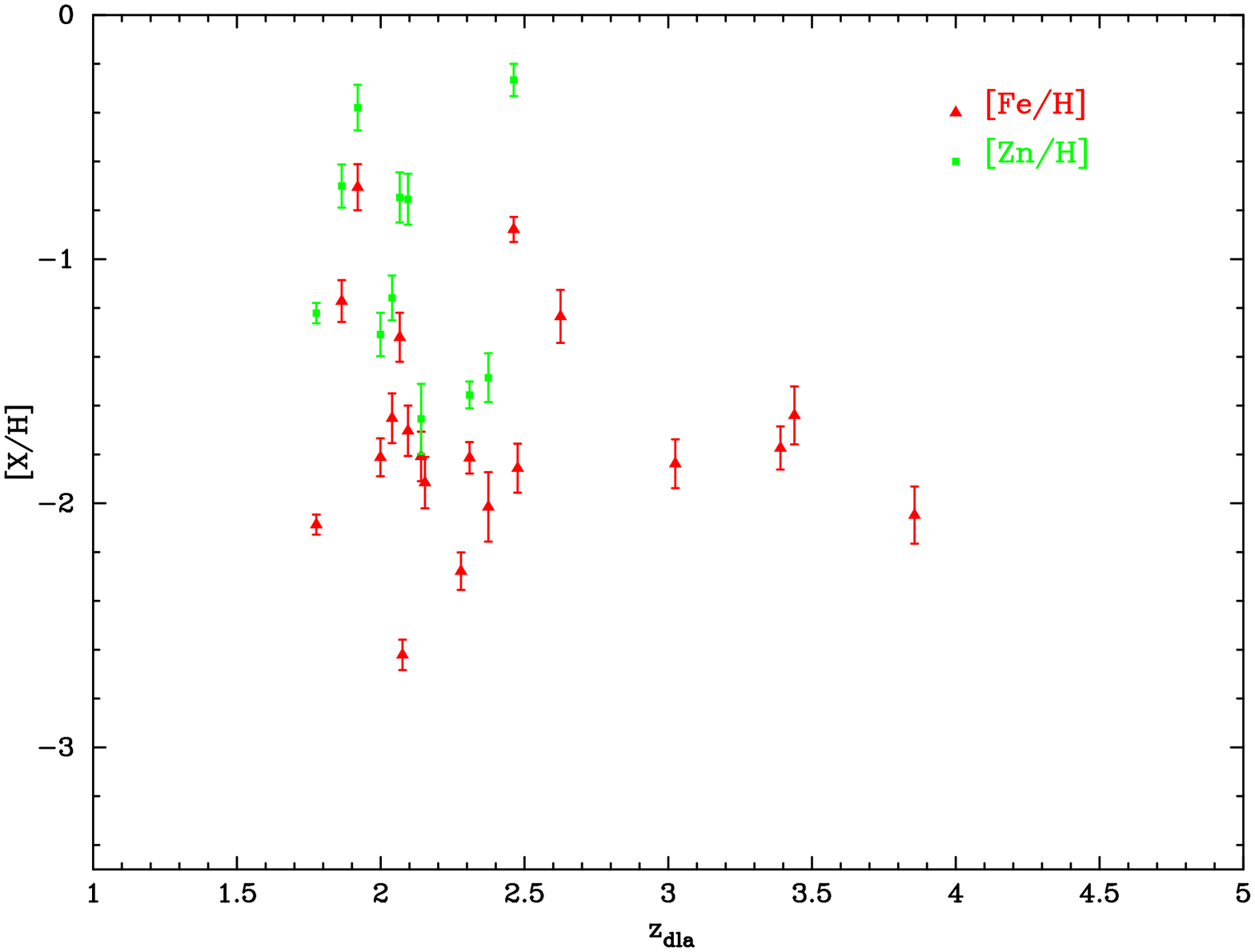}
\caption{ [Zn/H] and [Fe/H] measurements from the full sample of damped
\lya systems vs. redshift. }
\label{fig-metll}
\end{center}
\end{figure}

\section{METALLICITY}
\label{sec-met}

We now turn to examine the metallicity of our sample of damped 
systems.  Given the debate on the presence of dust in the
damped \lya systems, we will consider both Zn/H and Fe/H.
Figure~\ref{fig-metll} plots our complete sample of [Zn/H] and
[Fe/H] measurements vs. redshift.  The column density-weighted
mean for Zn,

\begin{equation}
[<{\rm Zn/H}>] = \log {\sum_i \N{Zn^+}_i \over \sum_i
\N{H^0}_i} - \log ({\rm Zn/H})_\odot \cmma
\end{equation}

\noindent for our full sample is [$<$Zn/H$>$] = $-1.15 \pm 0.15$~dex\footnote{
Note the error estimate reflects the errors in the individual [Zn/H]
measurements and not the size of the data sample.}.
This result confirms that of Pettini et al.\ (1997).
For Fe, we find [$<$Fe/H$>$] = $-1.64 \pm 0.11$~dex 
for $z < 3$ and $-1.77 \pm 0.11$ dex for $z >3$.
Lu et al.\ (1997)
have used similar measurements (in fact several of the values presented
here)
to conclude that $z \approx 3$ marks the onset of significant star
formation in the damped \lya systems.  Their interpretation is based
on the fact that the damped systems exhibit a break in 
[$<$Fe/H$>$] at $z \approx 3$.
Formally,
our data does not support their conclusion, but the fact that
the high redshift [$<$Fe/H$>$] value is dominated by only three
systems (Q0000$-$26, Q0019$-$15, and Q0347$-$38) suggests our
result is probably suffering from small number statistics.

\section{SUMMARY AND CONCLUSIONS}
\label{sec-sum}

We have presented accurate ionic column density and abundance
measurements for \ndla damped \lya systems observed with HIRES
on the 10m W.M. Keck Telescope.  Throughout the paper we have
utilized the apparent column density techniques to analyze
the damped \lya profiles and have adopted $\N{HI}$ values from
the literature.  The main results of the paper are summarized as
follows:

\begin{enumerate}

\item The abundance patterns of our \ndla systems match those observed by
Lu et al.\ (1996). 
Therefore, our analysis confirms their primary
conclusion that the damped \lya systems exhibit abundance 
patterns representative of Type II SN enrichment with the major exception
of [Zn/Fe] and to a lesser extent [Ni/Fe].  The Zn and Ni patterns,
however, are in accordance with what one would expect for dust
depletion based on observations of the lightly depleted,
'warm' HI clouds in the
ISM.  While the combination of dust depletion and Type II SN enrichment fits
the majority of the observations, this interpretation
is ruled out by the observed values of [S/Fe].

\item  A majority of the damped \lya
elemental abundances are consistent
with a dust depletion pattern on top of an underlying solar abundance
pattern.  
Observations of Titanium and Manganese, however, strongly contradict
this interpretation.  
In every system where Ti is observed, we measure [Ti/Fe] $\geq 0$~dex 
consistent with
the observed overabundance found in metal-poor halo stars and therefore
suggestive of Type II SN enrichment.
Similarly, the observed underabundance of [Mn/Fe] (\cite{lu96b})
is opposite to the effects of dust depletion and 
therefore requires a nucleosynthetic explanation, albeit not
necessarily Type II SN yields.

\item Our metallicity measurements confirm the principal
results from the surveys
of Pettini and collaborators.  Specifically,
we find:  [$<$Zn/H$>$] = $-1.15 \pm 0.15$~dex,
[$<$Fe/H$>$] = $-1.64 \pm 0.11$~dex 
for $z < 3$, and $-1.77 \pm 0.11$ dex for $z >3$.
Although we do not observe an evolution in the column density-weighted 
Fe abundance with redshift -- as claimed by Lu et al.\ (1997) --
we expect this inconsistency lies
in the small number statistics of our high $z$ sample.

\item  For a number of damped \lya systems in our sample (e.g.\
Q1331, Q0201, Q2348),  we observe metal-line systems within
500 $\mkms$ of the damped system.  In the case of Q2348, for
example, a metal-line system exhibiting SiII, SiIV and AlIII
transitions is located only $100 \mkms$ from the strongest damped
\lya component.  The absence of FeII and OI absorption and the
SiII/SiIV ratio for this component, however, indicates 
this system is significantly ionized.  We believe the same is true
for the majority
of these neighboring metal-line systems (the system at $z=1.858$
toward Q2230$+$02 is a notable exception).
If they were identified independently of the damped system,
these systems would be very strong Lyman limit systems.
Their coincidence with the damped system suggests they lie
within the halo enclosing the damped system or perhaps that
of a neighboring protogalactic system.  We expect a detailed analysis of
these systems may provide important insight
into the physical conditions surrounding the damped \lya systems.

\end{enumerate}

\acknowledgements

We acknowledge very helpful discussions with A. McWilliam, M. Pettini,
and G. Fuller.
We also would like to thank L. Storrie-Lombardi and I. Hook
for helping to provide target lists and $\N{HI}$ measurements.
We acknowledge the very helpful Keck support staff for their efforts
in performing these observations.
AMW and JXP were partially supported by 
NASA grant NAGW-2119 and NSF grant AST 86-9420443.

\setcounter{figure}{1}

\begin{figure}
\caption{Velocity plot of the metal-line transitions for the 
damped \lya system at $z = 3.439$ toward Q0019$-$15.
The vertical line at $v=0$ corresponds to $z = 3.43866$.}
\label{vpQ0019}
\end{figure}

\begin{figure}
\caption{Velocity plot of the metal-line transitions for the 
damped \lya system at $z = 2.309$ toward Q0100$+$13.
The vertical line at $v=0$ corresponds to $z = 2.309$.}
\label{vpQ0100}
\end{figure}

\begin{figure}
\caption{Velocity plot of the metal-line transitions for the 
damped \lya system at $z = 2.141$ toward Q0149$+$33.
The vertical line at $v=0$ corresponds to $z = 2.140755$.}
\label{vpQ0149}
\end{figure}

\begin{figure}
\caption{Velocity plot of the metal-line transitions for the 
damped \lya system at $z = 3.025$ toward Q0347$-$38.
The vertical line at $v=0$ corresponds to $z = 3.0247$.}
\label{vpQ0347}
\end{figure}

\begin{figure}
\caption{Velocity plot of the metal-line transitions for the 
damped \lya system at $z = 2.040$ toward Q0458$-$02.
The vertical line at $v=0$ corresponds to $z = 2.03955$.}
\label{vpQ0458}
\end{figure}

\begin{figure}
\caption{Velocity plot of the metal-line transitions for the 
damped \lya system at $z = 2.374$ toward Q0841$+$12.
The vertical line at $v=0$ corresponds to $z = 2.374518$.}
\label{vpQ0841A}
\end{figure}

\begin{figure}
\caption{Velocity plot of the metal-line transitions for the 
damped \lya system at $z = 2.476$ toward Q0841$+$12.
The vertical line at $v=0$ corresponds to $z = 2.476219$.}
\label{vpQ0841B}
\end{figure}

\begin{figure}
\caption{Velocity plot of the metal-line transitions for the 
damped \lya system at $z = 3.386$ toward Q0951$-$04.
The vertical line at $v=0$ corresponds to $z = 3.856689$.}
\label{vpQ0951A}
\end{figure}

\begin{figure}
\caption{Velocity plot of the metal-line transitions for the 
damped \lya system at $z = 4.203$ toward Q0951$-$04.
The vertical line at $v=0$ corresponds to $z = 4.202896$.}
\label{vpQ0951B}
\end{figure}

\begin{figure}
\caption{Velocity plot of the metal-line transitions for the 
damped \lya system at $z = 1.999$ toward Q1215$+$33.
The vertical line at $v=0$ corresponds to $z = 1.9991$.}
\label{vpQ1215}
\end{figure}

\begin{figure}
\caption{Velocity plot of the metal-line transitions for the 
damped \lya system at $z = 1.776$ toward Q1331$+$17.
The vertical line at $v=0$ corresponds to $z = 1.77636$.}
\label{vpQ1331}
\end{figure}

\begin{figure}
\caption{Velocity plot of the metal-line transitions for the 
damped \lya system at $z = 3.736$ toward Q1346$-$03.
The vertical line at $v=0$ corresponds to $z = 3.735830$.}
\label{vpQ1346}
\end{figure}

\begin{figure}
\caption{Velocity plot of the metal-line transitions for the 
damped \lya system at $z = 2.625$ toward Q1759$+$75.
The vertical line at $v=0$ corresponds to $z = 2.6253$.}
\label{vpQ1759}
\end{figure}

\begin{figure}
\caption{Velocity plot of the metal-line transitions for the 
damped \lya system at $z = 1.864$ toward Q2230$+$02.
The vertical line at $v=0$ corresponds to $z = 1.864388$.}
\label{vpQ2230}
\end{figure}

\clearpage

\begin{figure}
\caption{Velocity plot of the metal-line transitions for the 
presumed damped system at $z = 1.859$ toward Q2230$+$02.
The vertical line at $v=0$ corresponds to $z = 1.858536$.}
\label{vpQ2230B}
\end{figure}

\begin{figure}
\caption{Velocity plot of the metal-line transitions for the 
damped \lya system at $z = 2.066$ toward Q2231$-$00.
The vertical line at $v=0$ corresponds to $z = 2.06615$.}
\label{vpQ2231}
\end{figure}

\begin{figure}
\caption{Velocity plot of the metal-line transitions for the 
damped \lya system at $z = 2.279$ toward Q2348$-$14.
The vertical line at $v=0$ corresponds to $z = 2.2794$.}
\label{vpQ2348B}
\end{figure}

\begin{figure}
\caption{\lya profile for the damped system toward Q2348$-$14.
The overplotted curves correspond to $\N{HI} = 20.56 \pm 0.075$.} 
\label{Q2348-lya}
\end{figure}

\begin{figure}
\caption{Velocity plot of the metal-line transitions for the 
damped \lya system at $z = 2.095$ toward Q2359$-$02.
The vertical line at $v=0$ corresponds to $z = 2.095067$.}
\label{vpQ2359A}
\end{figure}

\begin{figure}
\caption{Velocity plot of the metal-line transitions for the 
damped \lya system at $z = 2.154$ toward Q2359$-$02.
The vertical line at $v=0$ corresponds to $z = 2.153934$.}
\label{vpQ2359B}
\end{figure}


\clearpage


\begin{thebibliography}{}

\bibitem[Evardsson et al.\ 1993]{evd93}
Evardsson, B., Anderson, J., Gutasfsson, B., Lambert, D.L.,
Nissen, P.E., and Tompkin, J. 1993, Astronomy and Astrophysics, 275, 101.

\bibitem[Fall \& Pei 1993]{fal93}
Fall, S.M. \& Pei, Y.C. 1993, \apj, 402, 479

\bibitem[Ferland 1991]{fer91}
Ferland, G. J. 1991, Ohio State Internal Report 91-01

\bibitem[Hoffman et al.\ 1996]{hff96}
Hoffman, R.D. et al.\ 1996, \apj, 460, 478

\bibitem[Howk \& Savage 1998]{howk98}
Howk, J.C. \& Savage, B.D. 1998, \apj, submitted

\bibitem[Kulkarni et al.\ 1997]{kulk97}
Kulkarni, V.P., Fall, S.M., \& Truran, J.W. 1997, \apj, 484, 7

\bibitem[Lipman \& Pettini 1995]{lpp95}  	
Lipman, K. \& Pettini, M. 1995, \apj, 442, 628

\bibitem[Lu et al.\ 1996]{lu96b}
Lu, L., Sargent, W.L.W., Barlow, T.A.,
Churchill, C.W., \& Vogt, S. 1996, \apjsupp, 107, 475

\bibitem[Lu et al.\ 1997]{lu97}
Lu, L., Sargent, W.L.W., Barlow, T.A.,
1997, 

\bibitem[Malaney and Chaboyer 1996]{mny96}
Malaney, R.A. and Chaboyer, B. 1996, \apj, 462, 57

\bibitem[McWilliam 1997]{mcwll97}
McWilliam, Andrew 1997, \araa, 35, 503

\bibitem[Morton 1991]{mor91}
Morton, D.C. 1991, \apjsupp, 77, 119

\bibitem[Nakamura et al.\ 1998]{nak98}
Nakamura, T., Umeda, H., Nomoto, K., Thielemann, F.,
\& Burrows, A. 1998, \apj, submitted (astro-ph/9809307)

\bibitem[Pettini et al.\ 1994]{ptt94}
Pettini, M., Smith, L. J., Hunstead, R. W., and King,
D. L. 1994, \apj, 426, 79

\bibitem[Pettini et al.\ 1995]{ptt95}
Pettini, M., Lipman, K., \& Hunstead, R.W. 1995, \apj, 451, 100

\bibitem[Pettini et al.\ 1997]{ptt97}
Pettini, M., Smith, L.J., King, D.L., \& Hunstead, R.W. 1997,
\apj, 486, 665

\bibitem[Pettini et al.\ 1998]{ptt98}
Pettini, M., Ellison, S., Steidel, C.C., \& Bowen, D.V. 1998,
\apj, in press

\bibitem[Prochaska \& Wolfe 1996]{pro96}
Prochaska, J. X. \& Wolfe, A. M. 1996, \apj, 470, 403

\bibitem[Prochaska \& Wolfe 1997a]{pro97a}
Prochaska, J. X. \& Wolfe, A. M. 1997, \apj, 474, 140

\bibitem[Prochaska \& Wolfe 1997b]{pro97b}
Prochaska, J. X. \& Wolfe, A. M. 1997, \apj, 486, 73

\bibitem[Prochaska \& Wolfe 1998]{pro98}
Prochaska, J. X. \& Wolfe, A. M. 1998, \apj, in press

\bibitem[Savage and Sembach 1991]{sav91}
Savage, B. D. and Sembach, K. R. 1991, \apj, 379, 245

\bibitem[Savage and Sembach 1996]{sav96} 
Savage, B. D. and Sembach, K. R. 1996, ARA\&A, 34, 279

\bibitem[Songaila et al.\ 1994]{song94}
Songaila et al.\ 1994, \nat, 371, 43

\bibitem[Sneden et al.\ 1991]{sne91}
Sneden, C., Gratton, R.G., \& Crocker, D.A. 1991,
A \& A, 246, 354

\bibitem[Storrie-Lombardi and Wolfe 1998]{storr98}
Storrie-Lombardi, L.J. \& Wolfe, A.M. 1998, in preparation

\bibitem[Tripp et al.\ 1996]{trp96}
Tripp, T. M., Lu L., \& Savage B.D. 1996, \apjsupp, 102, 239

\bibitem[Viegas 1994]{vgs94}
Viegas, S.M. 1994, \mnras, 276, 268

\bibitem[Vladilo 1998]{vld98}
Vladilo, G. 1998, \apj, 493, 583

\bibitem[Vogt 1992]{vgt92}
Vogt, S. S. 1992, in {\em ESO Conf. and Workshop Proc. 40,
High Resolution Spectroscopy with the VLT}, ed. M.-H. Ulrich (Garching:
ESO), p. 223

\bibitem[Welty et al.\ 1997]{wlty97}
Welty, D.E., Lauroesch, J.T., Blades, J.C., Hobbs, L.M.,
\& York, D.G. 1997, \apj, 489, 672

\bibitem[Wolfe \& Davis 1979]{wol79}
Wolfe, A.M., \& Davis, M.M. 1979, \aj, 84, 699

\bibitem[Wolfe et al.\ 1986]{wol86}
Wolfe, A.M., Turnshek, D.A., Smith, H.E., \& Cohen, R.D.
1986, \apjs, 61, 249

\bibitem[Wolfe et al.\ 1995]{wol95}
Wolfe, A. M., Lanzetta, K. M., Foltz, C. B., and
Chaffee, F. H. 1995, \apj, 454, 698

\bibitem[Wolfe \& Prochaska 1998]{wol98}
Wolfe, A.M. \& Prochaska, J.X. 1998, \apj, 494, 15L
    
\bibitem[Zsarg$\rm \acute o$ \& Federman 1998]{zsrgo98}
Zsarg$\rm \acute o$, J. \& Federman, S.R. 1998, \apj, 498, 256


\end{thebibliography}
\end{document}